\documentclass[sigconf]{acmart}

\usepackage{amsfonts,amssymb}

\newcommand{\ignore}[1]{}
\usepackage{xspace}
\newcommand{\paratitle}[1]{\vspace{1.5ex}\noindent\textbf{#1}}
\newcommand{\ie}{\emph{i.e.,}\xspace}

\newcommand{\eg}{\emph{e.g.,}\xspace}

\newcommand{\be}{\mathbf{e}}

\newcommand{\bb}{\mathbf{b}}

\newcommand{\bE}{\mathbf{E}}
\newcommand{\bF}{\mathbf{F}}
\newcommand{\bW}{\mathbf{W}}

\newcommand{\bM}{\mathbf{M}}

\newcommand{\bP}{\mathbf{P}}
\newcommand{\bX}{\mathbf{X}}

\newcommand{\bK}{\mathbf{K}}

\usepackage{multirow}
\usepackage{multicol}
\usepackage{caption}
\usepackage{subcaption}
\usepackage{latexsym}
\usepackage{mflogo}
\usepackage{graphics}

\AtBeginDocument{%
  \providecommand\BibTeX{{%
    \normalfont B\kern-0.5em{\scshape i\kern-0.25em b}\kern-0.8em\TeX}}}

\copyrightyear{2022}
\acmYear{2022}
\setcopyright{acmcopyright}
\acmConference[WWW '22] {Proceedings of the ACM Web Conference 2022}{April 25--29, 2022}{Virtual Event, Lyon, France.}
\acmBooktitle{Proceedings of the ACM Web Conference 2022 (WWW '22), April 25--29, 2022, Virtual Event, Lyon, France}
\acmPrice{15.00}
\acmISBN{978-1-4503-9096-5/22/04}
\acmDOI{10.1145/3485447.3512111}

\begin{document}




\title{Filter-enhanced MLP is All You Need for \\
Sequential Recommendation}

\author{Kun Zhou$^{1,4\dagger}$, Hui Yu$^{2,5\dagger}$, Wayne Xin Zhao$^{3,4^*}$ and Ji-Rong Wen$^{3,4}$}\thanks{$^\dagger$Equal contribution} \thanks{$^*$Corresponding author.}
\affiliation{%
\institution{$^1$School of Information, Renmin University of China\country{China}}
\institution{$^2$University of Chinese Academy of Sciences\country{China}}
\institution{$^3$Gaoling School of Artificial Intelligence, Renmin University of China\country{China}}
\institution{$^4$Beijing Key Laboratory of Big Data Management and Analysis Methods}
\institution{$^5$Key Laboratory of Petroleum Resources Research, Institute of Geology and Geophysics, Chinese Academy of Sciences}
}
\affiliation{%
  \institution{francis\_kun\_zhou@163.com, ishyu@outlook.com, batmanfly@gmail.com, jrwen@ruc.edu.cn}
}
\renewcommand{\shortauthors}{Kun Zhou, Hui Yu, Wayne Xin Zhao, \& Ji-Rong Wen}



\begin{abstract}
Recently,  deep neural networks such as RNN, CNN and Transformer have been applied in the task of  sequential recommendation, which aims to capture the dynamic preference characteristics from logged user behavior data for accurate recommendation.
However, in online platforms, logged user behavior data is inevitable to contain noise, and  deep recommendation models are easy to overfit on these logged data.
To tackle this problem, we borrow the idea of filtering algorithms from  signal processing that attenuates the noise in the frequency domain.
In our empirical experiments, we find that filtering algorithms can substantially improve representative sequential recommendation models, and integrating simple filtering algorithms (\eg Band-Stop Filter) with an all-MLP architecture can even outperform competitive Transformer-based models.
Motivated by it, we propose \textbf{FMLP-Rec}, an all-MLP model with learnable filters for sequential recommendation task.
The all-MLP architecture endows our model with lower time complexity, and the learnable filters can adaptively attenuate the noise information in the frequency domain.
Extensive experiments conducted on eight real-world datasets demonstrate the superiority of our proposed method over competitive RNN, CNN, GNN and Transformer-based methods.
Our code and data are publicly available at the link: \textcolor{blue}{\url{https://github.com/RUCAIBox/FMLP-Rec}}.
\end{abstract}




\maketitle
\section{Introduction}
Recommender systems~\cite{lu2015recommender,zhang2019deep,zhou2020improving,DBLP:journals/corr/abs-2011-01731} have been widely deployed in online platforms (\eg Amazon and Taobao) for predicting the potential interests of users over a large item pool.
In real-world applications, users' behaviors are dynamic and evolving over time. Thus, it is critical to capture the sequential characteristics of  user behaviors for making appropriate recommendations, which is the core goal for sequential recommendation~\cite{DBLP:journals/corr/HidasiKBT15,DBLP:conf/icdm/KangM18,DBLP:conf/www/RendleFS10,DBLP:conf/wsdm/TangW18}.

To characterize the evolving patterns of users' historical behaviors, a number of sequential recommendation models have been developed in the literature based on deep neural networks~\cite{zhang2019deep}.
Typical solutions are based on RNN~\cite{DBLP:journals/corr/HidasiKBT15,DBLP:conf/icdm/LiuWWLW16} and CNN~\cite{DBLP:conf/wsdm/TangW18,DBLP:conf/wsdm/YuanKAJ019}.
Furthermore, a surge of works adopt more advanced neural network architectures (\eg memory network~\cite{DBLP:conf/sigir/HuangZDWC18,DBLP:conf/wsdm/ChenXZT0QZ18} and self-attention mechanism~\cite{DBLP:conf/cikm/LiRCRLM17,DBLP:conf/wsdm/LiWM20}) to enhance the modeling capacity for effectively capturing dynamic user preference.
More recently, Transformer-based approaches~\cite{DBLP:conf/icdm/KangM18,DBLP:conf/cikm/SunLWPLOJ19,DBLP:conf/cikm/ZhouWZZWZWW20} have shown remarkable performance in this task by stacking multi-head self-attention layers. 

However, stacked self-attention layers involve a large number of parameters, which might lead to the over-parameterized architecture of Transformer-based methods~\cite{fan2019reducing,mehta2020delight}.
Besides, these methods mainly fit the model parameters based on logged user behavior data, which are in essence noisy~\cite{agichtein2006improving,said2012users} or even contains malicious fakes~\cite{DBLP:conf/www/ZhangLD020,DBLP:conf/cikm/ChenL19}. 
It has been found that deep neural networks tend to overfit on noisy data~\cite{caruana2001overfitting,lever2016points}. The case becomes more severe for self-attention based recommenders when the logged sequence data contains noise, since it attends to all items for sequence modeling. 

Considering the above issues, we aim to simplify the Transformer-based sequential recommender as well as increase its robustness to resist the noise in logged data. 
Our key idea is borrowed from the digital signal processing field, where filtering algorithms are used to reduce the influence of noise~\cite{rabiner1975theory} for sequence data.
We suspect that when the sequence data was denoised, it would become easier to capture sequential user behaviors. If this was true, we might be able to simplify the heavy self-attention components from Transformer-based approaches. To examine this hypothesis, we conduct several empirical experiments in Section~\ref{sec-motivation} by simply denoising the item embeddings with three classical filtering algorithms (more details are referred to Table~\ref{tab:motivation}). We find that filtering algorithms can substantially improve these deep sequential recommendation models, including the Transformer-based SASRec~\cite{DBLP:conf/icdm/KangM18}.

Motivated by the empirical findings, we propose a novel \textbf{F}ilter-enhanced \textbf{MLP} approach for sequential \textbf{Rec}ommendation, named \textbf{FMLP-Rec}. By removing the self-attention components from Transformers, FMLP-Rec is solely based on MLP structures for stacking blocks.
As the major technical contribution, we incorporate a filter component in each stacked block, where we perform Fast Fourier Transform~(FFT)~\cite{soliman1990continuous} to convert the input representations into the frequency domain and an inverse FFT procedure recovers the denoised representations.
The filter component plays a key role in reducing the influence of the noise from item representations\footnote{Note that we do not learn to remove the noisy items, since it is a more difficult task in practice, usually without any  ground-truth labels. Instead, we consider directly improving the item representations in order to reduce the potential influence of noise.}. To implement it, we incorporate learnable filters to encode the input item sequences in the frequency domain, which can be optimized from the raw data without human priors. 
Our approach can effectively attenuate noise information and extract meaningful features from all the frequencies (\eg long/short-term item interactions).
Theoretically speaking, according to convolution theorem~\cite{soliman1990continuous}, it can be proved that learnable filters are equivalent to the circular convolution in the time domain, which has a larger receptive field on the whole sequence, and can better capture periodic characteristics of user behaviors. Another merit of the filter component is that it requires less time cost without considering pairwise item correlations, which results in a lighter and faster network architecture. 

To the best of our knowledge, it is the first time that a filter-enhanced all-MLP architecture has been applied to the sequential recommendation task, which is simple, effective and efficient.
To validate the effectiveness of our model, we conduct extensive experiments on eight real-world datasets from different scenarios for sequential recommendations. Experimental results show that FMLP-Rec outperforms state-of-the-art RNN, CNN, GNN and Transformer-based baseline models.

\ignore{
Our main contributions are summarized as follows:
(1) It is the first time that a filter-enhanced all-MLP architecture has been applied to the sequential recommendation task, which is simple, effective and efficient.
(2) Our approach incorporates learnable filters to alleviate the influence of noise in user behavior sequences. The learnable filters are equivalent to the circular convolutions that have larger receptive fields and can capture periodic characteristics.
(3) Extensive experiments conducted on eight real-world datasets demonstrate the effectiveness of our proposed approach.}

\section{PRELIMINARIES}
In this section, we formulate the sequential recommendation problem and then introduce the discrete Fourier transform.

\subsection{Problem Statement}
Assume that we have a set of users and items, denoted by $\mathcal{U}$ and $\mathcal{I}$, respectively, where $u\in \mathcal{U}$ denotes a user and $i\in \mathcal{I}$ denotes an item. The numbers of users and items are denoted as $|\mathcal{U}|$ and $|\mathcal{I}|$, respectively. 
For sequential recommendation with implicit feedback, a user $u$ has a context $c$, a chronologically-ordered interaction sequence with items: $c=\{i_1,\cdots,i_n\}$, where $n$ is the number of interactions and $i_t$ is the $t$-th item that the user $u$ has interacted with. For convenience, we use $i_{j:k}$ to denote the subsequence, \ie $i_{j:k}=\{{i_{j},\cdots,i_{k}}\}$ where $1\leq j<k\leq n$.

Based on the above notations, we now define the task of sequential recommendation. Formally, given the contextual item sequence of a user $c=\{i_1,\cdots,i_n\}$, the task of sequential recommendation is to predict the next item that the user is likely to interact with at the $(n+1)$-th step, denoted as $p(i_{n+1}|i_{1:n})$. 

\subsection{Fourier Transform}
\paratitle{Discrete Fourier transform.}
Discrete Fourier transform (DFT) is essential in the digital signal processing area~\cite{soliman1990continuous,rabiner1975theory} and is a crucial component in our approach. In this paper, we only consider the 1D DFT.
Given a sequence of numbers $\{x_{n}\}$ with $n\in [0,N-1]$, the 1D DFT converts the sequence into the frequency domain by:
\begin{equation}
\small
    \label{eq:dft}
    X_k = \sum_{n=0}^{N-1} x_n e^{-{\frac{2\pi i}{N}} n k}, \quad 0 \leq k \leq N-1.
\end{equation}
where $i$ is the imaginary unit. For each $k$, the DFT generates a new representation $X_k$ as a sum of all the original input tokens $x_n$ with so-called ``twiddle factors''. In this way, $X_k$ represents the spectrum of the sequence $\{x_n\}$ at the frequency $\omega_k = 2\pi k/N$. 
Note that DFT is an one-to-one transformation. Given the DFT $X_k$, we can recover the original sequence $\{x_{n}\}$ by the inverse DFT (IDFT):
\begin{equation}
\small
    x_n = \frac{1}{N} \sum_{k=0}^{N - 1}X_k e^{\frac{2\pi i}{N} nk}.
    \label{equ:idft}
\end{equation}

\paratitle{Fast Fourier Transform.}
To compute the DFT, the Fast Fourier Transform (FFT) is widely used in previous works~\cite{heideman1985gauss,van1992computational}. The standard FFT algorithm is the Cooley–Tukey algorithm \cite{cooley1965algorithm, frigo2005design}, which recursively re-expresses the DFT of a sequence of length $N$ and reduces the time complexity to $O(N \log N)$.
The inverse DFT in Eq.~\ref{equ:idft}, which has a similar form to the DFT, can also be computed efficiently using the inverse fast Fourier transform (IFFT).

Since FFT can convert input signals into the frequency domain where the periodic characteristics are easier to capture, it is widely used in digital signal processing area to filter noise signals~\cite{soliman1990continuous,staelin1969fast,anderson1984model}.
A commonly-used way is the Low-Pass Filter (LPF) that attenuates high-frequency noise signals after processed by FFT.
In this paper, we consider using FFT and filtering algorithms to reduce the influence of noisy features within the user interacted item sequence.
\section{Empirical Analysis with Filtering Algorithms for Recommendation}
\label{sec-motivation}
In this section, we conduct an empirical study to test:
(1)  the effectiveness of filtering algorithms in sequential recommendation models, and (2) the effectiveness of integrating   filtering algorithms with all-MLP architectures.

\subsection{Analysis Setup}
\label{sec-setup}
For empirical study, we select the Amazon~\cite{DBLP:conf/sigir/McAuleyTSH15} \emph{Beauty} and \emph{Sports} datasets for the evaluation of sequential recommendation methods.

\paratitle{Sequential recommendation algorithms}. We conduct experiments on GRU4Rec~\cite{DBLP:journals/corr/HidasiKBT15} and SASRec~\cite{DBLP:conf/icdm/KangM18}, two representative sequential recommendation models.
The two models largely follow the standard framework of deep sequential models~\cite{DBLP:conf/icdm/Rendle10,DBLP:journals/corr/HidasiKBT15,DBLP:conf/wsdm/TangW18}, consisting of an embedding layer, a sequence encoder layer and a prediction layer, but adopt RNN and Transformer in the sequence encoder layer, respectively.
Despite that the two models have shown promising results, they may not be robust to noise in the user behavior sequence~\cite{DBLP:conf/ijcai/SunWSY21,DBLP:conf/recsys/YueHZM21}.
Thus, we directly add a non-parameter filter layer between the embedding layer and the sequence encoder layer of the two models, and do not change other components.

\paratitle{Filtering algorithms}. In the filter layer, given the embedding matrix of the item sequence, we conduct the following operations for each dimension of features: \emph{FFT} $\rightarrow$ \emph{Filtering Algorithm} $\rightarrow$ \emph{IFFT}.
After filtering, we take the denoised embedding matrix as the input of the sequence encoder layer.
For filtering algorithms, we select three classical methods~\cite{soliman1990continuous} as follows:

$\bullet$ \emph{High-Pass Filter (HPF)} passes signals with a higher frequency and attenuates ones with a lower frequency. After FFT, we set the values of the lower-frequency half of signals into zero.

$\bullet$ \emph{Low-Pass Filter (LPF)} passes signals with a lower frequency and attenuates ones with a higher frequency. After FFT, we set the values of the higher-frequency half of signals into zero.

$\bullet$ \emph{Band-Stop Filter (BSF)} attenuates signals with a medium frequency, and passes others. After FFT, we set the values of the medium-frequency half of signals into zero.

\subsection{Results and Findings}
\label{sec-filter-algo}


\begin{table}[t]
	\small
	\centering
	\caption{Performance comparison of SASRec and GRU4Rec with different filtering algorithms.}
	\label{tab:motivation}
	\setlength{\tabcolsep}{1.3mm}{
		\begin{tabular}{l|l||cc|cc}
			\hline
			\multirow{2}*{Methods} & \multirow{2}*{Filter} &  \multicolumn{2}{c|}{Beauty} &
			\multicolumn{2}{c}{Sports} \\
			\cline{3-6}
			& & HR@10 & NDCG@10 & HR@10 &NDCG@10 \\
			\cline{1-6}
			\multirow{5} * {GRU4Rec}
			& & 0.4106 & 0.2584 &0.4299 &0.2527 \\
			&+HPF  & 0.3828 & 0.2228 & 0.3654 & 0.2063\\
			&+LPF  & 0.4351 & \textbf{0.2689} & \textbf{0.4481} & \textbf{0.2578}\\
			&+BSF  & \textbf{0.4372} & 0.2658 & 0.4432 & 0.2563\\
			\cline{1-6}
			\multirow{5} * {SASRec}
			&  &0.4696 &0.3156 &0.4622 &0.2869 \\
			&+HPF  &0.4544 &0.3037 &0.4530 &0.2785\\
			&+LPF  &0.4941 &0.3320 &0.5040 &0.3138\\
			&+BSF  &\textbf{0.5011} &\textbf{0.3334} &\textbf{0.5115} &\textbf{0.3172}\\
			\hline
		\end{tabular}
	}
\end{table}

In this part, we present the results and discuss the findings. 

\paratitle{The effect on representative models}. 
We report the results in Table~\ref{tab:motivation}.
As can be seen, adding Low-Pass Filter (LPF) and Band-Stop Filter (BSF) lead to consistent improvements, especially on SASRec.
For GRU4Rec, LPF achieves the best performance than other filtering algorithms, while for SASRec, BSF achieves the best.
In contrast, High-Pass Filter (HPF) causes performance degradation in most cases. 
From these observations, we can conclude that:

$\bullet$ The item embedding matrix are likely to contains noise that affects the performance of sequential recommendation models.

$\bullet$  A proper filtering algorithm on the embedding layer is useful to alleviate the above problem. But for different models, the most suitable filtering algorithm may be also different.

$\bullet$  The low-frequency information within the embedding matrix seems more important for sequential recommendation. This phenomenon is similar to findings in hydrology~\cite{sang2009relation}, seismology~\cite{stammler1993seismichandler} and praxiology~\cite{unuma1995fourier} that low-frequency signals in nature and human behavior are usually meaningful periodic characteristics.

\begin{figure}[t!]
    \centering
    \includegraphics[width=\linewidth]{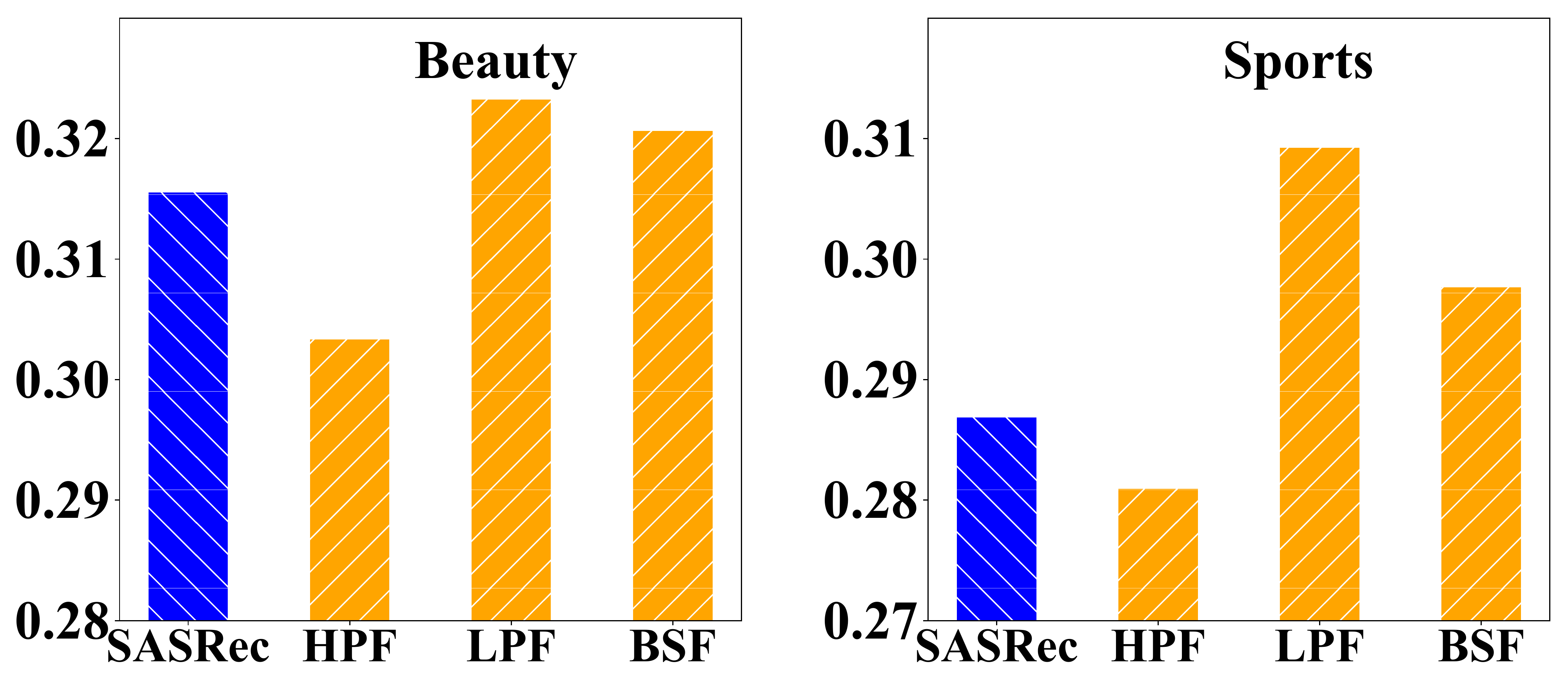}
    \caption{Performance (NDCG@10) comparison of all-MLP variants of SASRec with different filtering algorithms.}
    \label{fig-motivation}
\end{figure}

\paratitle{The effect on all-MLP models}.
The above study confirms that filtering algorithms can improve the performance of RNN and Transformer based sequential recommendation models.
In this part, we continue to examine the performance of  a simply variant 
that integrates these filtering algorithms with all-MLP architectures.
Based on the architecture of SASRec~\cite{DBLP:conf/icdm/KangM18}, we remove the multi-head self-attention blocks within the Transformer-based sequence encoder layer, but add a filter layer after the embedding layer.
We also select HPF, LPF and BSF algorithms as in section~\ref{sec-setup}, and other components are not changed. 
In this way, the  variant models only rely on MLPs to model the item sequence.

We report the performance of the all-MLP variant models with SASRec in Figure~\ref{fig-motivation}. 
As we can see, after removing multi-head self-attention blocks, most of the model variants still perform well. 
And the variant model with LPF even outperforms SASRec model with a large margin. 
It indicates that proper filtering algorithms can inspire the potential of simple all-MLP models to surpass complex Transformer-based models.
By removing both the noise information and the self-attention blocks, the model more lightweight, which reduces the 
 risk of overfitting.
Based on the above analysis, it is promising to design an effective and efficient all-MLP model with a proper filtering algorithm for sequential recommendation.
\section{Method}
The empirical findings in Section~\ref{sec-motivation} have demonstrated that an all-MLP architecture with proper filtering algorithms (\eg LPF) can yield very good recommendation performance.
However, previous filtering algorithms~\cite{soliman1990continuous} usually require domain knowledge or expert efforts in designing proper filters and setting hyperparameters (\eg filtering thresholds). It is still a challenge to effectively integrate the filtering algorithms into existing deep recommendation frameworks. 
In this section, we present an all-MLP architecture (named as \textbf{FMLP-Rec}) for sequential recommendation by stacking MLP blocks with learnable filters, which can automatically learn proper filters for various sequential recommendation scenarios. 


\begin{figure}
\includegraphics[width=.96\linewidth]{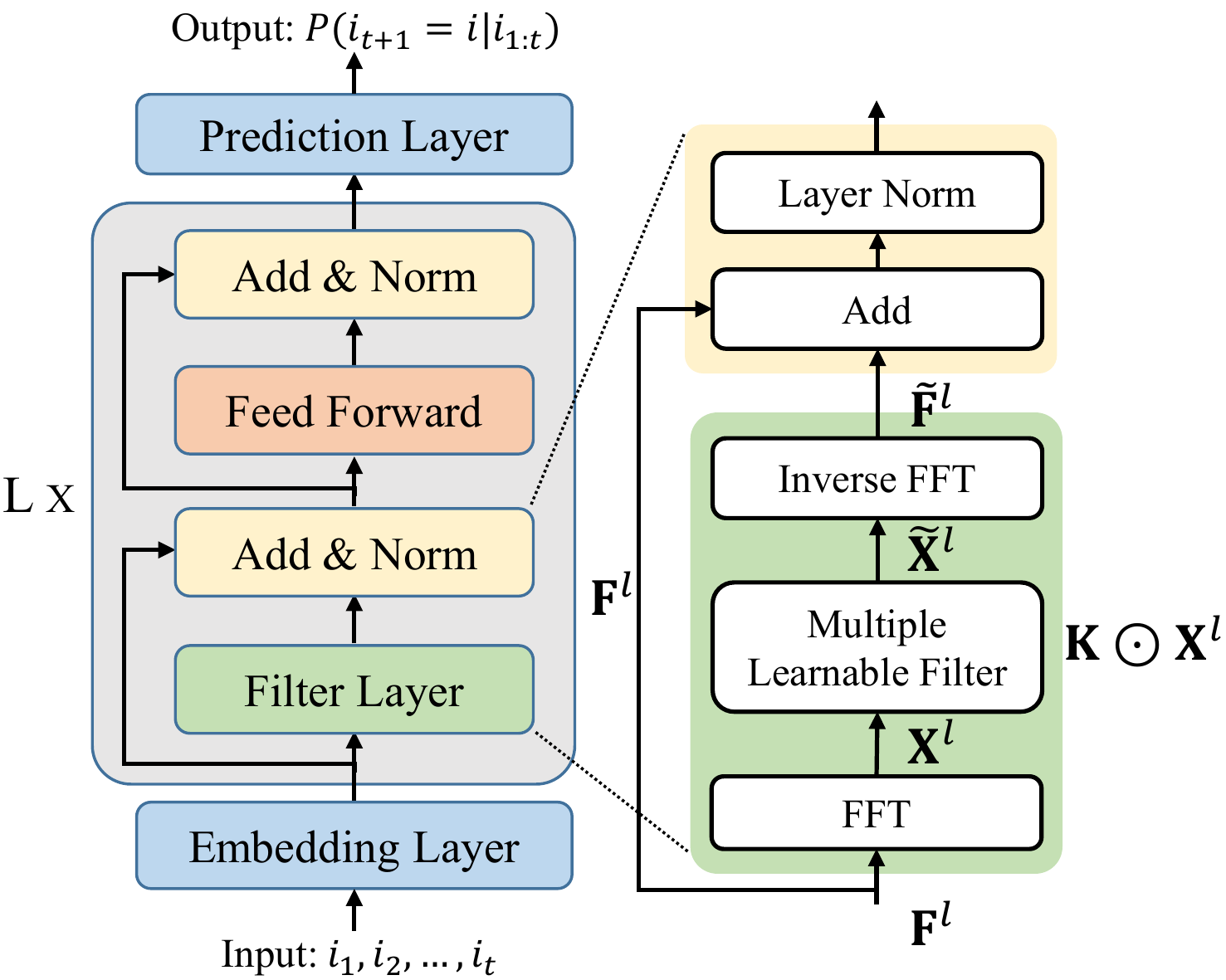}
\caption{The overview of our FMLP-Rec, an all-MLP model that stacks multiple learnable filter-enhanced blocks.}
\label{approach}
\end{figure}

\subsection{FMLP-Rec: An All-MLP Sequential Recommender with Learnable Filters}
Similar to the original Transformer architecture, our FMLP-Rec also stacks multiple neural blocks to produce the representation of sequential user preference for recommendation. 
The key difference of our approach is to replace the multi-head self-attention structure in Transformer with a novel filter structure.
Besides the effect of noise attenuation by filters, such a structure is mathematically equivalent to the circular convolution (proved in Section~\ref{sec-rcnn}), which can also capture sequence-level preference characteristics~\cite{soliman1990continuous}. 
Next, we present the details of our approach.



\subsubsection{Embedding Layer}
In the embedding layer, we maintain an item embedding matrix $\bM_{I}\in\mathbb{R}^{|\mathcal{I}|\times d}$ to project the high-dimensional one-hot representation of an item to low-dimensional dense representation.
Given a $n$-length item sequence, we apply a look-up operation from $\bM_{I}$ to form the input embedding matrix $\bE\in \mathbb{R}^{n\times d}$. 
Besides, we incorporate a learnable position encoding matrix $\bP\in \mathbb{R}^{n\times d}$ to enhance the input representation of the item sequence. By this means, the sequence representation $\bE_{I}\in \mathbb{R}^{n\times d}$ can be obtained by summing the two embedding matrices.
Since the item and position embedding matrices are randomly initialized, it may affect the filtering mechanism and cause the training process unstable.
Inspired by recent works~\cite{DBLP:conf/icdm/KangM18,DBLP:conf/naacl/DevlinCLT19}, we perform dropout and the layer normalization operations to alleviate these problems. Thus, we generate the sequence representation $\bE_{I}\in \mathbb{R}^{n\times d}$ by:
\begin{equation}
    \bE_{I}=Dropout(LayerNorm(\bE+\bP)).
\end{equation}

\subsubsection{Learnable Filter-enhanced Blocks}
Based on the embedding layer, we develop the item encoder by stacking multiple learnable filter blocks. 
A learnable filter block generally consists of two sub-layers, \ie a filter layer and a point-wise feed-forward network. 

\paratitle{Filter Layer}. In the filter layer, we perform filtering operation for each dimension of features in the frequency domain, and then perform skip connection and layer normalization.
Given the input item representation matrix $\bF^{l}\in \mathbb{R}^{n\times d}$ of the $l$-th layer (when $l=0$, we set $\bF^{0}=\bE_{I}$), we first perform FFT along the item dimension to convert $\bF^{l}$ to the frequency domain:
\begin{equation}
    \mathbf{X}^{l} = \mathcal{F}(\bF^{l}) \in \mathbb{C}^{n\times d}
\end{equation}
where $\mathcal{F}(\cdot)$ denotes the one-dimensional FFT. Note that $\mathbf{X}^{l}$ is a complex tensor and represents the spectrum of $\bF^{l}$. We can then modulate the spectrum by multiplying a learnable filter $\bW\in \mathbb{C}^{n\times d}$:
\begin{equation}
    \widetilde{\mathbf{X}}^{l} = \bW\odot \mathbf{X}^{l},
    \label{eq-multi-fre}
\end{equation}
where $\odot$ is the element-wise multiplication. The filter $\bK$ is called the \emph{learnable filter} since it can be optimized by SGD to adaptively represent an arbitrary filter in the frequency domain. Finally, we adopt the inverse FFT to transform the modulated spectrum $\widetilde{\bX}^{l}$ back to the time domain and update the sequence representations:
\begin{equation}
    \widetilde{\mathbf{F}}^{l}\leftarrow \mathcal{F}^{-1}(\widetilde{\mathbf{X}}^{l}) \in \mathbb{R}^{n\times d}.
\end{equation}
where $\mathcal{F}^{-1}(\cdot)$ denotes the inverse 1D FFT, which converts the complex tensor into a real number tensor. 
With the operations of FFT and inverse FFT, the noise from the logged data can be effectively reduced, and we can therefore obtain purer item embeddings. 
Following SASRec~\cite{DBLP:conf/icdm/KangM18}, we also incorporate the skip connection~\cite{DBLP:conf/cvpr/HeZRS16}, layer normalization~\cite{DBLP:journals/corr/BaKH16} and dropout~\cite{srivastava2014dropout} operations to alleviate the gradient vanishing and unstable training problems as:
\begin{equation}
\label{eq-ln+drop}
    \widetilde{\bF}^{l} = \text{LayerNorm}(\bF^{l} + \text{Dropout}(\widetilde{\bF}^{l}))
\end{equation}

\paratitle{Feed-forward layers}. 
In the point-wise feed-forward network, we incorporate MLP and ReLU activation functions to further capture the non-linearity characteristics.
The computation is defined as:
\begin{align}
    \text{FFN}(\widetilde{\bF}^{l}) &= \big(\text{ReLU}(\widetilde{\bF}^{l}\bW_{1}+\bb_{1})\big)\bW_{2}+\bb_{2},
\end{align}
where $\bW_{1}$, $\bb_{1}$, $\bW_{2}$, $\bb_{2}$ are trainable parameters. Then, we also perform skip connection and layer normalization operations as in Eq.~\ref{eq-ln+drop} to generate the output of the $l$-layer.

\subsubsection{Prediction Layer}
In the final layer of FMLP-Rec, we calculate the user's preference score for the item $i$ in step $(t+1)$ under the context from user history as:
\begin{eqnarray}
    P(i_{t+1}=i|i_{1:t})=\be_{i}^{\top}\bF^{L}_{t},
\end{eqnarray}
where $\be_{i}$ is the representation of item $i$ from item embedding matrix $\bM_{I}$,
$\bF_{t}^{L}$ is the output of the $L$-layer learnable filter blocks at  step $t$, and $L$ is the number of learnable filter blocks.
We adopt the pairwise rank loss to optimize the model parameters as:
\begin{align}
\label{ft}
    L = -\sum_{u \in \mathcal{U}}\sum_{t=1}^{n}\log \sigma\bigg(P(i_{t+1}|i_{1:t})-P(i_{t+1}^{-}|i_{1:t})\bigg),
\end{align}
where we pair each ground-truth item $i_{t+1}$ with a negative item $i_{t+1}^{-}$ that is randomly sampled.

\subsection{Theoretical Analysis with Filter Layers}
\label{sec-rcnn}
Besides noise attenuation, we now show that the proposed filter blocks can also capture sequential characteristics from logged data. 
We first theoretically prove that our proposed learnable filter is equivalent to circular convolution.

In the filter layer of FMLP-Rec,  the input information is firstly converted to spectrum representations in the frequency domain via FFT, and then further multiplied the learnable filter $\bW$.  
In this way, the formulation of the learnable filter $\bW$ can be regarded as a set of learnable frequency filters $\{\mathbf{w}^{(1)}, \cdots, \mathbf{w}^{(d)}\}$ for different hidden dimensions in the item embedding matrix, where $d$ denotes the hidden size.
According to the convolution theorem~\cite{rabiner1975theory,soliman1990continuous}, the \emph{multiplication} in the frequency domain is equivalent to the \emph{circular convolution} in the time domain.
It is a special case of periodic convolution between two periodic functions that have the same period.
For the $t$-th dimension features $\{f^{(t)}_n\}_{n=0}^{N-1}$ from the item representation matrix $\bF^{l}$ and a filter $\{h^{(t)}_n\}_{n=0}^{N-1}=\mathcal{F}^{-1}(w^{(t)})$, their circular convolution is defined as $\{y^{(t)}_n\}_{n=0}^{N-1}$, where
\begin{equation}
    y^{(t)}_n = \sum_{m=0}^{N-1} h^{(t)}_{m} \cdot f^{(t)}_{(n-m) \text{~mod~} N}
\end{equation}
where $\text{mod}$  denotes the integer modulo operation and $n$ is the sequence length.
Consider that the DFT of the sequence $\{y^{(t)}_n\}_{n=0}^{N-1}$ is $\{\widetilde{x}^{(t)}_n\}_{n=0}^{N-1}$, we have the following derivations:
\begin{equation}
\begin{split}
 \widetilde{x}^{(t)}_k &= \sum_{n=0}^{N-1} \sum_{m=0}^{N-1}h^{(t)}_{m}f^{(t)}_{(n-m)\%N}e^{-\frac{2\pi i}{N}kn}\\
 &=\sum_{m=0}^{N-1}h^{(t)}_{m}e^{-\frac{2\pi i}{N}km}\sum_{n=0}^{N-1}f^{(t)}_{(n-m)\%N}e^{-\frac{2\pi i}{N}k(n-m)}\\
 &=w^{(t)}_{k}\left(\sum_{n=m}^{N-1}f^{(t)}_{n-m}e^{-\frac{2\pi i}{N}k(n-m)} + \sum_{n=0}^{m-1}f^{(t)}_{n-m + N}e^{-\frac{2\pi i}{N}k(n-m)}\right)\\
 &=w^{(t)}_{k}\left(\sum_{n=0}^{N-m-1}f^{(t)}_{n}e^{-\frac{2\pi i}{N}kn} + \sum_{n=N-m}^{N-1}f^{(t)}_{n}e^{-\frac{2\pi i}{N}kn}\right)\\
 &=w^{(t)}_{k}\sum_{n=0}^{N-1}f^{(t)}_{n}e^{-\frac{2\pi i}{N}kn}=w^{(t)}_{k}x^{(t)}_{k}, \nonumber
\end{split}
\end{equation}
where in the right hand, $x^{(t)}_{k}$ is exactly the $k$-th number of the $t$-th dimension feature from $\bX$ in the frequency domain, and $w^{(t)}_{k}$ is the $k$-th weight from the $t$-th filter $w^{(t)}$.
In conclusion, we prove that 
\begin{equation}
    f^{(t)}*h^{(t)}=\mathcal{F}^{-1}(\mathbf{w}^{(t)}\odot \mathbf{x}^{(t)})
\end{equation}
where ``$*$'' and ``$\odot$'' denotes circular convolution and element-wise multiplication, respectively.
Therefore, the multiplication of the $t$-th dimension feature from item representations $\bX^{l}$ and the $t$-th filter $\mathbf{w}^{(t)}$ from $\mathbf{W}$ in Eq.~\ref{eq-multi-fre}, is equivalent to the circular convolution operation on the $t$-th feature of $\bF^{l}$ using convolution kernel $\{h^{(t)}_n\}_{n=0}^{N-1}$.
Similar to RNN, CNN and Transformer, the proposed filter block is also able to capture sequential characteristics, since it has in essence the same effect of circular convolution.
Besides, compared with traditional linear convolution in previous works~\cite{DBLP:conf/wsdm/TangW18,DBLP:conf/wsdm/YuanKAJ019}, the circular convolution has a larger receptive field on the entire sequence and is able to better capture \emph{periodic patterns}~\cite{soliman1990continuous}.
Such merit is particularly appealing for the recommendation task, where users' behaviors tend to show certain periodic trends~\cite{hsu2007consumer,jin2013understanding,crespo2010influence}.

\subsection{Discussion}

\subsubsection{Comparison with Transformer-based Sequential Recommenders}
Transformer-based models such as SASRec~\cite{DBLP:conf/icdm/KangM18} and BERT4Rec~\cite{DBLP:conf/cikm/SunLWPLOJ19} typically stack multi-head self-attention blocks for learning the sequential representations.
It relies on a heavy self-attention structure to learn item-to-item correlations. 
As a comparison, our approach FMLP-Rec directly removes all the self-attention structures, and the entire approach is solely based on MLP-based structures by integrating additional filter layers. Such a design can largely reduce both the space and time complexity for modeling sequence data.
More importantly, we have shown that the learnable filter layer is equivalent to the circular convolution operation using convolution kernels with the same size as the feature map~\cite{rabiner1975theory,soliman1990continuous}.
As a result, it can own the same receptive field as the self-attention mechanism but largely reduces the number of involved parameters. 
Besides, the circular convolution is able to capture periodic characteristics, which is also an important feature for sequential recommendation.


\ignore{\subsubsection{Computational Complexity Analysis}
Compared with state-of-the-art Transformer-based model, our proposed MLP-Rec is also more efficient.
Since the self-attention layers within Transformer-based models requires to calculate the similarity for each item pairs. Its time complexity is $\mathcal{O}(n^{2})$~\cite{DBLP:conf/nips/VaswaniSPUJGKP17,DBLP:journals/corr/abs-2105-14103}, where $n$ denotes the sequence length.
In contrast, our MLP-Rec only consists of FFT operations $\mathcal{O}(n\log n)$~\cite{heideman1985gauss,van1992computational}, point-wise feed-forward networks $\mathcal{O}(n)$ and an Inverse FFT operation $\mathcal{O}(n\log n)$, which means the total time complexity is $\mathcal{O}(n\log n)$.
Besides, the item-item similarity matrix in Transformer-based models also cause $\mathcal{O}(n^2)$ space complexity.
As a comparison, the space complexity of the FFT, IFFT and MLP layer in MLP-Rec is $\mathcal{O}(n)$.
Therefore, our MLP-Rec owns both lower time complexity and lower space complexity than Transformer-based models.}

\subsubsection{Time Complexity and Receptive Field Analysis}
\begin{table}[t]
	\small
	\centering
	\caption{Time complexity and receptive field of the proposed FMLP-Rec with Caser and SASRec, where $n$ and $k$ denote the sequence length and convolution kernel size, respectively. For simplicity, we omit other same terms (\eg hidden size) and highlight the difference at  the sequence level.}
	\label{tab:complexity}
	\setlength{\tabcolsep}{1.3mm}{
		\begin{tabular}{l|c|c}
			\hline
			& \textbf{Time Complexity per Layer} & \textbf{Receptive Field} \\
			\hline
			Caser & $\mathcal{O}(kn)$ & $k$ \\
			SASRec & $\mathcal{O}(n^2)$ & $n$ \\
			FMLP-Rec & $\mathcal{O}(n\log n)$ & $n$ \\
			\hline
		\end{tabular}
	}
\end{table}

In this part, we compare our model with  representative sequential recommendation models in terms of time complexity and receptive field.  
We select CNN-based Caser~\cite{DBLP:conf/wsdm/TangW18} and Transformer-based SASRec~\cite{DBLP:conf/icdm/KangM18} as comparisons, since they represent two lines of different neural architectures. 
Since these models involve different algorithmic details (\eg dropout and normalization), it may not be fair to directly compare them. 
Instead, we consider a more general comparison by only on the sequence length $n$.
The comparison results of the model complexity and receptive field are shown in Table~\ref{tab:complexity}.

First, Caser performs convolution operations in time domains, so its time complexity and receptive field are $\mathcal{O}(k\cdot n)$ and $k$, respectively.
However, since Caser relies on convolution kernels to capture sequential pattern characteristics, it usually requires a larger kernel size to extend the receptive field for better performance.
Besides, in Transformer-based models, the self-attention layers require calculating the similarity for each pair of items. 
Its time complexity and receptive field are $\mathcal{O}(n^{2})$ and $n$~\cite{DBLP:conf/nips/VaswaniSPUJGKP17,DBLP:journals/corr/abs-2105-14103}, respectively.
In contrast, our FMLP-Rec consists of FFT and IFFT operations with the time cost of $\mathcal{O}(n\log n)$~\cite{heideman1985gauss,van1992computational}, and point-wise feed-forward networks with the time cost of $\mathcal{O}(n)$, which means the total time complexity is $\mathcal{O}(n\log n)$.
Besides, since our filter layer is equivalent to the circular convolution on the whole sequence, its receptive field is the same as Transformer.
Therefore, our FMLP-Rec can achieve larger receptive field and meanwhile lower complexity.

\section{Experiment}

\begin{table}
    \small
	\caption{Statistics of the datasets after preprocessing.}
	\label{tab:datasets}
	\begin{tabular}{l|r|r|r|r}
	\toprule
	\textbf{Dataset} & \# \textbf{Sequences} & \# \textbf{Items} & \# \textbf{Actions} & \# \textbf{Sparsity} \\
	\midrule
	\midrule
	Beauty & 22,363 & 12,101 & 198,502 & 99.93\% \\
	Sports & 25,598 & 18,357 & 296,337 & 99.95\% \\
	Toys & 19,412 & 11,924 & 167,597 & 99.93\% \\
	Yelp & 30,431 & 20,033 & 316,354 & 99.95\% \\
	Nowplaying & 145,612 & 59,593 & 1,085,410 & 99.99\% \\
	Retailrocket & 321,032 & 51,428 & 871,637 & 99.99\% \\
	Tmall & 66,909 & 37,367 & 427,797 & 99.98\% \\
	Yoochoose & 470,477 & 19,690 & 1,434,349 & 99.98\%\\
	\bottomrule
	\end{tabular}
\end{table}

\subsection{Experimental Setup}
\subsubsection{Dataset}
We conduct experiments on eight datasets with varying domains. Their statistics are summarized in Table~\ref{tab:datasets}.

(1) \textbf{Beauty, Sports, and Toys}: these three datasets are obtained from Amazon review datasets in \cite{DBLP:conf/sigir/McAuleyTSH15}. We select three subcategories: Beauty, Sports and Outdoors, and Toys and Games.

(2) \textbf{Yelp}\footnote{\url{https://www.yelp.com/dataset}} is a dataset for business recommendation. As it is very large, we only use the transaction records after \textit{January 1st, 2019}.

(3) \textbf{Nowplaying}\footnote{\url{https://dbis.uibk.ac.at/node/263\#nowplaying}} contains music listening events collected from Twitter, where users posted tracks that were currently listening. 

(4) \textbf{RetailRocket}\footnote{\url{https://www.kaggle.com/retailrocket/ecommerce-dataset}} is collected from a personalized e-commerce website. It contains six months of user browsing activities.

(5) \textbf{Tmall}\footnote{\url{https://tianchi.aliyun.com/dataset/dataDetail?dataId=42}} comes from IJCAI-15 competition, which contains user’s shopping logs on Tmall online shopping platform.

(6) \textbf{Yoochoose}\footnote{\url{https://www.kaggle.com/chadgostopp/recsys-challenge-2015}} contains a collection of sessions from a retailer, where each session encapsulates the click events.

Note that the item sequences in Beauty, Sports, Toys and Yelp are user transaction records, while in Nowplaying, RetailRocket, Tmall and Yoochoose are click sessions.
For all datasets, we group the interaction records by users or sessions, and sort them by the timestamps ascendingly. Following~\cite{DBLP:conf/www/RendleFS10,DBLP:conf/recsys/HidasiQKT16}, we filter unpopular items and inactive users with fewer than five interaction records.

\subsubsection{Evaluation Metrics}
Following \cite{zhao2020revisiting}, we employ top-$k$ Hit Ratio (HR@$k$), top-$k$ Normalized Discounted Cumulative Gain (NDCG@$k$), and Mean Reciprocal Rank (MRR) for evaluation, which are widely used in related works~\cite{DBLP:conf/www/RendleFS10,DBLP:conf/icdm/KangM18,DBLP:conf/cikm/ZhouWZZWZWW20}.
Since HR@1 is equal to NDCG@1, we report results on HR@\{1, 5, 10\}, NGCG@\{5, 10\}, and MRR.
Following the common strategy~\cite{DBLP:conf/sigir/HuangZDWC18, DBLP:conf/icdm/KangM18}, we pair the ground-truth item with 99 randomly sampled negative items that the user has not interacted with. We calculate all metrics according to the ranking of the items and report the average score.
Note that we also rank the ground-truth item with all candidate items and report the full-ranking results in supplementary materials.

\subsubsection{Baseline Models}
\begin{table*}[t!]
    \small
	\caption{Performance comparison of different methods on four datasets containing user transaction records. The best performance and the second best performance methods are denoted in bold and underlined fonts respectively.}
	\label{tab:main_table_cikm2020}
	\setlength{\tabcolsep}{1.3mm}{
	\begin{tabular}{llccccccccccccc}
	\toprule
		Datasets & Metric & PopRec &FM& AutoInt &GRU4Rec & Caser &HGN &RepeatNet&CLEA &SASRec &BERT4Rec &SRGNN &GCSAN &FMLP-Rec  \\
	\midrule
\multirow{6} * {Beauty}
 &HR@1   &0.0678 &0.0405 &0.0447 &0.1337 &0.1337  &0.1683 &0.1578 &0.1325 &0.1870 &0.1531 &0.1729 &\underline{0.1973} &\textbf{0.2011} \\
 &HR@5   &0.2105 &0.1461 &0.1705 &0.3125 &0.3032  &0.3544 &0.3268 &0.3305 &\underline{0.3741} &0.3640 &0.3518 &0.3678 &\textbf{0.4025} \\
 &NDCG@5 &0.1391 &0.0934 &0.1063 &0.2268 &0.2219  &0.2656 &0.2455 &0.2353 &0.2848 &0.2622 &0.2660 &\underline{0.2864} &\textbf{0.3070} \\
 &HR@10  &0.3386 &0.2311 &0.2872 &0.4106 &0.3942  &0.4503 &0.4205 &0.4426 &0.4696 &\underline{0.4739} &0.4484 &0.4542 &\textbf{0.4998} \\
 &NDCG@10&0.1803 &0.1207 &0.1440 &0.2584 &0.2512  &0.2965 &0.2757 &0.2715 &\underline{0.3156} &0.2975 &0.2971 &0.3143 &\textbf{0.3385} \\
 &MRR    &0.1558 &0.1096 &0.1226 &0.2308 &0.2263  &0.2669 &0.2498 &0.2376 &0.2852 &0.2614 &0.2686 &\underline{0.2882} &\textbf{0.3051} \\
\hline
\multirow{6} * {Sports}
 &HR@1   &0.0763 &0.0489 &0.0644 &0.1160 &0.1135  &0.1428 &0.1334 &0.1114 &0.1455 &0.1255 &0.1419 &\textbf{0.1669} &\underline{0.1646} \\
 &HR@5   &0.2293 &0.1603 &0.1982 &0.3055 &0.2866  &0.3349 &0.3162 &0.3041 &0.3466 &0.3375 &0.3367 &\underline{0.3588} &\textbf{0.3803} \\
 &NDCG@5 &0.1538 &0.1048 &0.1316 &0.2126 &0.2020  &0.2420 &0.2274 &0.2096 &0.2497 &0.2341 &0.2418 &\underline{0.2658} &\textbf{0.2760} \\
 &HR@10  &0.3423 &0.2491 &0.2967 &0.4299 &0.4014  &0.4551 &0.4324 &0.4274 &0.4622 &0.4722 &0.4545 &\underline{0.4737} &\textbf{0.5059} \\
 &NDCG@10&0.1902 &0.1334 &0.1633 &0.2527 &0.2390  &0.2806 &0.2649 &0.2493 &0.2869 &0.2775 &0.2799 &\underline{0.3029} &\textbf{0.3165} \\
 &MRR    &0.1660 &0.1202 &0.1435 &0.2191 &0.2100  &0.2469 &0.2334 &0.2156 &0.2520 &0.2378 &0.2461 &\underline{0.2691} &\textbf{0.2763} \\
\hline
\multirow{6} * {Toys}
 &HR@1   &0.0585 &0.0257 &0.0448 &0.0997 &0.1114  &0.1504 &0.1333 &0.1104 &0.1878 &0.1262 &0.1600 &\textbf{0.1996} &\underline{0.1935} \\
 &HR@5   &0.1977 &0.0978 &0.1471 &0.2795 &0.2614  &0.3276 &0.3001 &0.3055 &\underline{0.3682} &0.3344 &0.3389 &0.3613 &\textbf{0.4063} \\
 &NDCG@5 &0.1286 &0.0614 &0.0960 &0.1919 &0.1885  &0.2423 &0.2192 &0.2102 &0.2820 &0.2327 &0.2528 &\underline{0.2836} &\textbf{0.3046} \\
 &HR@10  &0.3008 &0.1715 &0.2369 &0.3896 &0.3540  &0.4211 &0.4015 &0.4207 &\underline{0.4663} &0.4493 &0.4413 &0.4509 &\textbf{0.5062} \\
 &NDCG@10&0.1618 &0.0850 &0.1248 &0.2274 &0.2183  &0.2724 &0.2517 &0.2473 &\underline{0.3136} &0.2698 &0.2857 &0.3125 &\textbf{0.3368} \\
 &MRR    &0.1430 &0.0819 &0.1131 &0.1973 &0.1967  &0.2454 &0.2253 &0.2138 &0.2842 &0.2338 &0.2566 &0.\underline{2871} &\textbf{0.3012} \\
\hline
\multirow{6} * {Yelp}
 &HR@1   &0.0801 &0.0624 &0.0731 &0.2053 &0.2188  &0.2428 &0.2341 &0.2102 &0.2375 &0.2405 &0.2176 &\underline{0.2493} &\textbf{0.2727} \\
 &HR@5   &0.2415 &0.2036 &0.2249 &0.5437 &0.5111  &0.5768 &0.5357 &0.5707 &0.5745 &\underline{0.5976} &0.5442 &0.5725 &\textbf{0.6191} \\
 &NDCG@5 &0.1622 &0.1333 &0.1501 &0.3784 &0.3696  &0.4162 &0.3894 &0.3955 &0.4113 &\underline{0.4252} &0.3860 &0.4162 &\textbf{0.4527} \\
 &HR@10  &0.3609 &0.3153 &0.3367 &0.7265 &0.6661  &0.7411 &0.6897 &0.7473 &0.7373 &\underline{0.7597} &0.7096 &0.7371 &\textbf{0.7720} \\
 &NDCG@10&0.2007 &0.1692 &0.1860 &0.4375 &0.4198  &0.4695 &0.4393 &0.4527 &0.4642 &\underline{0.4778} &0.4395 &0.4696 &\textbf{0.5024} \\
 &MRR    &0.1740 &0.1470 &0.1616 &0.3630 &0.3595  &0.3988 &0.3769 &0.3751 &0.3927 &\underline{0.4026} &0.3711 &0.4006 &\textbf{0.4299} \\
\hline
\bottomrule
	\end{tabular}
	}
\end{table*}
We compare our proposed approach with the following  baseline methods:
(1) \textbf{PopRec} ranks items according to the popularity measured by the number of interactions;
(2) \textbf{FM}~\cite{DBLP:conf/icdm/Rendle10} characterizes the pairwise interactions between variables using factorized model;
(3) \textbf{AutoInt}~\cite{DBLP:conf/cikm/SongS0DX0T19} utilizes the multi-head self-attentive neural network to learn the feature interactions;
(4) \textbf{GRU4Rec}~\cite{DBLP:journals/corr/HidasiKBT15} applies GRU to model item sequences;
(5) \textbf{Caser}~\cite{DBLP:conf/wsdm/TangW18} is a CNN-based method that applies horizontal and vertical convolutions for sequential recommendation;
(6) \textbf{HGN}~\cite{DBLP:conf/kdd/MaKL19} adopts hierarchical gating networks to capture long-term and short-term user interests;
(7) \textbf{RepeatNet}~\cite{DBLP:conf/aaai/RenCLR0R19} adds a copy mechanism on RNN architecture that can choose items from a user’s history;
(8) \textbf{CLEA}~\cite{DBLP:conf/sigir/QinWL21} is recently proposed and performs item-level denoising via a contrastive learning model;
(9) \textbf{SASRec}~\cite{DBLP:conf/icdm/KangM18} is a unidirectional Transformer-based sequential recommendation model;
(10) \textbf{BERT4Rec}~\cite{DBLP:conf/cikm/SunLWPLOJ19} uses a Cloze objective loss for sequential recommendation by the bidirectional Transformer;
(11) \textbf{SRGNN}~\cite{DBLP:conf/aaai/WuT0WXT19} models session sequences as graph-structured data and uses an attention network;
(12) \textbf{GCSAN}~\cite{DBLP:conf/ijcai/XuZLSXZFZ19} utilizes both graph neural network and self-attention mechanism for session-based recommendation.

\subsection{Experimental Results}

\begin{table*}[t!]
    \small
	\caption{Performance comparison of different methods on four session-based datasets. Since these datasets do not have attribute information and the item sequences are usually shorter, we remove several improper baseline methods.}
	\label{tab:session_table_cikm2020}
	\setlength{\tabcolsep}{1.3mm}{
	\begin{tabular}{llcccccccccc}
	\toprule
		Datasets &Metric &PopRec &GRU4Rec &Caser &HGN &RepeatNet&CLEA &SASRec &SRGNN &GCSAN &FMLP-Rec \\
	\midrule
\multirow{6} * {Nowplaying}
&HR@1   &0.0757 &0.4035 &0.3435 &0.3491 &0.3350 &0.2728 &0.4396 &0.3819 &\underline{0.4447} &\textbf{0.4731} \\
&HR@5   &0.2197 &0.6829 &0.6267 &0.6026 &0.5257 &0.5575 &\underline{0.7042} &0.6028 &0.6728 &\textbf{0.7262} \\
&NDCG@5 &0.1480 &0.5536 &0.4942 &0.4835 &0.4355 &0.4226 &\underline{0.5812} &0.4986 &0.5658 &\textbf{0.6094} \\
&HR@10  &0.3318 &0.7720 &0.7318 &0.6992 &0.6110 &0.6766 &\underline{0.7968} &0.7007 &0.7616 &\textbf{0.8081} \\
&NDCG@10&0.1841 &0.5825 &0.5283 &0.5149 &0.4631 &0.4612 &\underline{0.6113} &0.5304 &0.5946 &\textbf{0.6360} \\
&MRR    &0.1602 &0.5314 &0.4752 &0.4677 &0.4297 &0.4070 &\underline{0.5615} &0.4892 &0.5520 &\textbf{0.5895} \\
\hline
\multirow{6} * {Retailrocket}
&HR@1   &0.0817 &0.7202 &0.6871 &0.6432 &0.6744 &0.3410 &0.7588 &0.7408 &\textbf{0.7773} &\underline{0.7736} \\
&HR@5   &0.2315 &0.8597 &0.8348 &0.7650 &0.7724 &0.6139 &\underline{0.8769} &0.8373 &0.8650 &\textbf{0.8810} \\
&NDCG@5 &0.1577 &0.7982 &0.7689 &0.7098 &0.7270 &0.4853 &0.8250 &0.7937 &\underline{0.8259} &\textbf{0.8338} \\
&HR@10  &0.3388 &0.8925 &0.8719 &0.8036 &0.8112 &0.7296 &\underline{0.9024} &0.8684 &0.8901 &\textbf{0.9060} \\
&NDCG@10&0.1922 &0.8089 &0.7809 &0.7223 &0.7395 &0.5227 &0.8333 &0.8037 &\underline{0.8340} &\textbf{0.8419} \\
&MRR    &0.1694 &0.7858 &0.7563 &0.7029 &0.7234 &0.4702 &0.8145 &0.7878 &\underline{0.8199} &\textbf{0.8247} \\
\hline
\multirow{6} * {Tmall}
 &HR@1  &0.1025 &0.4178 &0.3288 &0.4467 &\textbf{0.5556} &0.2895 &0.4045 &0.4026 &0.4650 &\underline{0.5173} \\
&HR@5   &0.2264 &0.5855 &0.5066 &0.5793 &\underline{0.6090} &0.4573 &0.5478 &0.5335 &0.5940 &\textbf{0.6565} \\
&NDCG@5 &0.1647 &0.5062 &0.4230 &0.5168 &\underline{0.5826} &0.3768 &0.4792 &0.4710 &0.5329 &\textbf{0.5911} \\
&HR@10  &0.2967 &\underline{0.6636} &0.5943 &0.6471 &0.6494 &0.5478 &0.6275 &0.6082 &0.6591 &\textbf{0.7206} \\
&NDCG@10&0.1874 &0.5315 &0.4513 &0.5386 &\underline{0.5956} &0.4060 &0.5049 &0.4950 &0.5538 &\textbf{0.6118} \\
&MRR    &0.1723 &0.5021 &0.4209 &0.5168 &\textbf{0.5894} &0.3775 &0.4804 &0.4736 &0.5328 &\underline{0.5879} \\
\hline
\multirow{6} * {Yoochoose}
 &HR@1  &0.1794 &0.7208 &0.7041 &0.6278 &0.7450 &0.5106 &0.7611 &0.7575 &\textbf{0.7855} &\underline{0.7749} \\
&HR@5   &0.4990 &0.8950 &0.8818 &0.8425 &0.8673 &0.7628 &0.8976 &0.8840 &\underline{0.9073} &\textbf{0.9084} \\
&NDCG@5 &0.3432 &0.8189 &0.8026 &0.7460 &0.8123 &0.6465 &0.8375 &0.8280 &\textbf{0.8535} &\underline{0.8499} \\
&HR@10  &0.6574 &0.9273 &0.9172 &0.8927 &0.9001 &0.8436 &0.9273 &0.9137 &\underline{0.9320} &\textbf{0.9359} \\
&NDCG@10&0.3948 &0.8294 &0.8141 &0.7623 &0.8230 &0.6727 &0.8471 &0.8376 &\textbf{0.8616} &\underline{0.8588} \\
&MRR    &0.3276 &0.8004 &0.7838 &0.7246 &0.8019 &0.6257 &0.8240 &0.8164 &\textbf{0.8414} &\underline{0.8364} \\
\bottomrule
	\end{tabular}
	}
\end{table*}

The results of different methods on datasets containing user transaction record are shown in Table~\ref{tab:main_table_cikm2020}, and results on session-based datasets are shown in Table~\ref{tab:session_table_cikm2020}. Based on the results, we can find:

First, non-sequential recommendation methods (\ie PopRec, FM and AutoInt) perform worse than sequential recommendation methods. It indicates that the sequential pattern is important in this task.
As for sequential recommendation methods, SASRec and BERT4Rec utilize Transformer-based architectures, and mostly achieve better performance than RNN-based models (\ie GRU4Rec and RepeatNet), CNN-based model (\ie Caser) and gate-based model (\ie HGN).
A possible reason is that Transformer-based models have more parameters corresponding to stronger  capacity to capture sequential characteristics.
Besides, CLEA achieves comparable performance with SASRec and BERT4Rec in part of datasets. 
Since CLEA adopts the item-level denoising strategy, it indicates that alleviating the influence of noise is useful to improve the recommendation performance. 
We can also see that GNN-based models (\ie SRGNN and GCSAN) also achieve comparable performance as Transformer-based models.
It suggests that GNNs are also promising to capture useful characteristics for accuracy recommendation.

Finally, by comparing our approach with all the baselines, it is clear to see that FMLP-Rec performs consistently better than them by a large margin on most of datasets. 
Different from these baselines, we adopt an all-MLP architecture with learnable filters to encode the item sequence.
The learnable filters can alleviate the influence of noise information, and are equivalent to circular convolutions that can capture the periodic characteristics in the item sequences with a larger receptive field.
The all-MLP architecture largely reduces the model scale, leading to a lower time complexity.
As a result, our FMLP-Rec is effective and efficient.
This result also shows that all-MLP architectures are effective for sequential recommendation.

\section{Further Analysis}

\begin{table}[t]
	\caption{Ablation  study of our FMLP-Rec, we report NDCG@10 on Beauty and Sports datasets.}
	\label{tab:filter}
	\setlength{\tabcolsep}{1.3mm}{
		\begin{tabular}{l|cc|cc}
			\hline
			 &  \multicolumn{2}{c|}{Beauty} &
			\multicolumn{2}{c}{Sports} \\
			\hline
			& HR@10 & NDCG@10 & HR@10 &NDCG@10 \\
			\hline
			\hline
			FMLP-Rec & \textbf{0.4998} & \textbf{0.3385} & \textbf{0.5059} & \textbf{0.3165}\\
			\hline
			w/o Filter Layer &0.4317 &0.2914 &0.4243 &0.2604\\
			w/o FFN & 0.4607 & 0.3067 & 0.4594 & 0.2838\\
			w/o Add \& Norm &0.4654 &0.2919 &0.4831 &0.2860 \\
			\hline
			+HPF & 0.4567 & 0.3034 & 0.4595 & 0.2810\\
			+LPF & 0.4847 & 0.3233 & 0.4949 & 0.3093\\
			+BSF & 0.4822 & 0.3207 & 0.4835 & 0.2977\\
			\hline
		\end{tabular}
	}
\end{table}

\subsection{Ablation tudy}

Our proposed FMLP-Rec contains filter layers, feed-forward network and Add \& Norm operations. 
To verify the effectiveness of each component, we conduct the ablation study on Beauty and Sports datasets to analyze the contribution of each part. 
Besides, to validate if the learnable filters are more useful than classical filtering algorithms, we also conduct variation study by replacing the learnable filters with high-pass, low-pass and band-stop filters.

From the results in Table~\ref{tab:filter}, we can observe that removing any components would lead to the performance degradation, especially the filter layer. It indicates all the components in FMLP-Rec are useful to improve the recommendation performance.
Besides, we can see that our FMLP-Rec outperforms all the other variants  with classical filtering algorithms.
The reason is that our learnable filters can adaptively learn to filter the noise information via SGD, which is promising to better adapt to the data distribution.

\subsection{Applying Learnable Filters to Other Models}
\begin{figure}[t]
    \centering
    \begin{subfigure}[b]{0.49\linewidth}
        \centering
        \includegraphics[width=\textwidth]{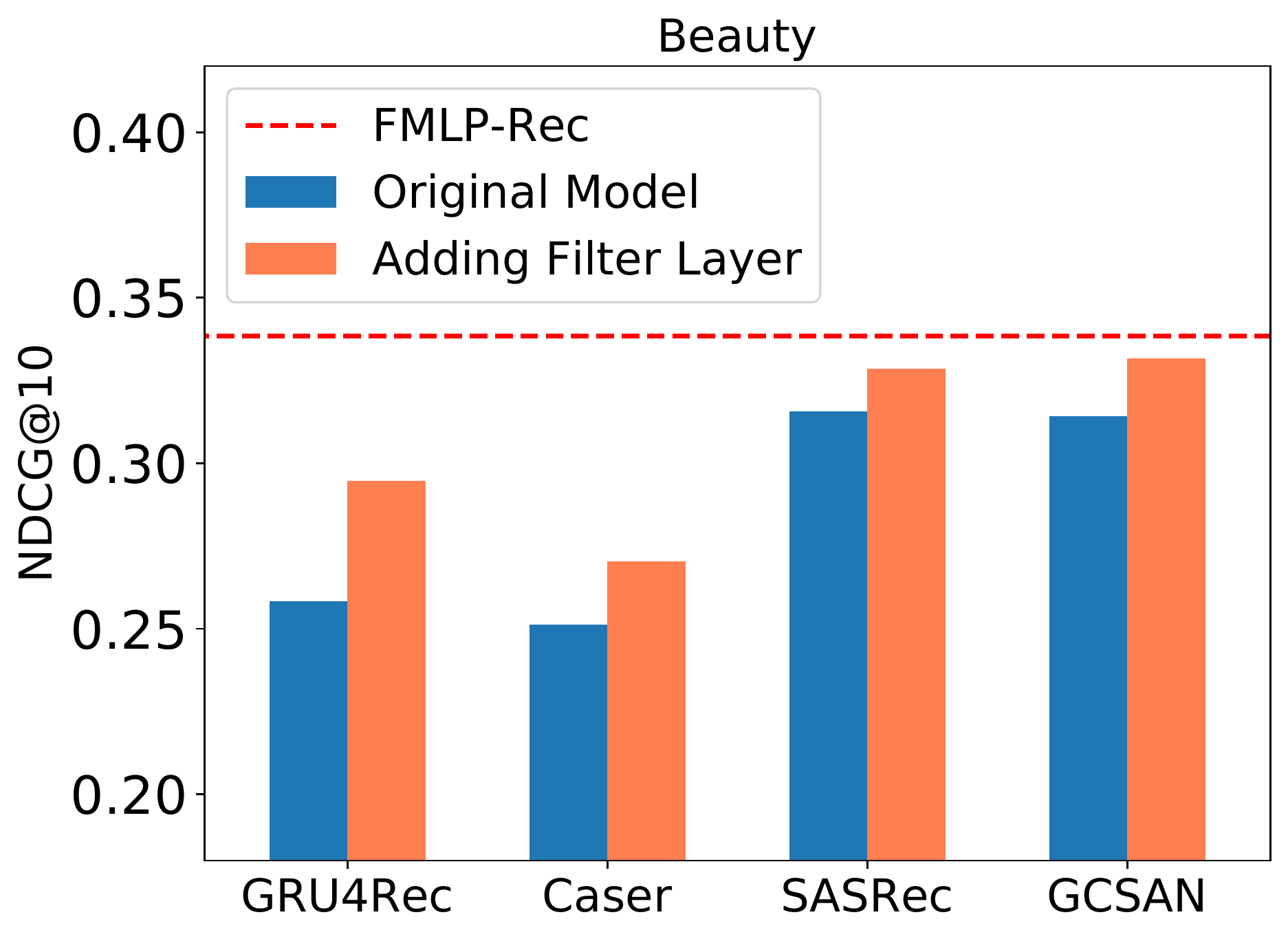}
        \label{beauty-filter}
    \end{subfigure}
    \begin{subfigure}[b]{0.49\linewidth}
        \centering
        \includegraphics[width=\textwidth]{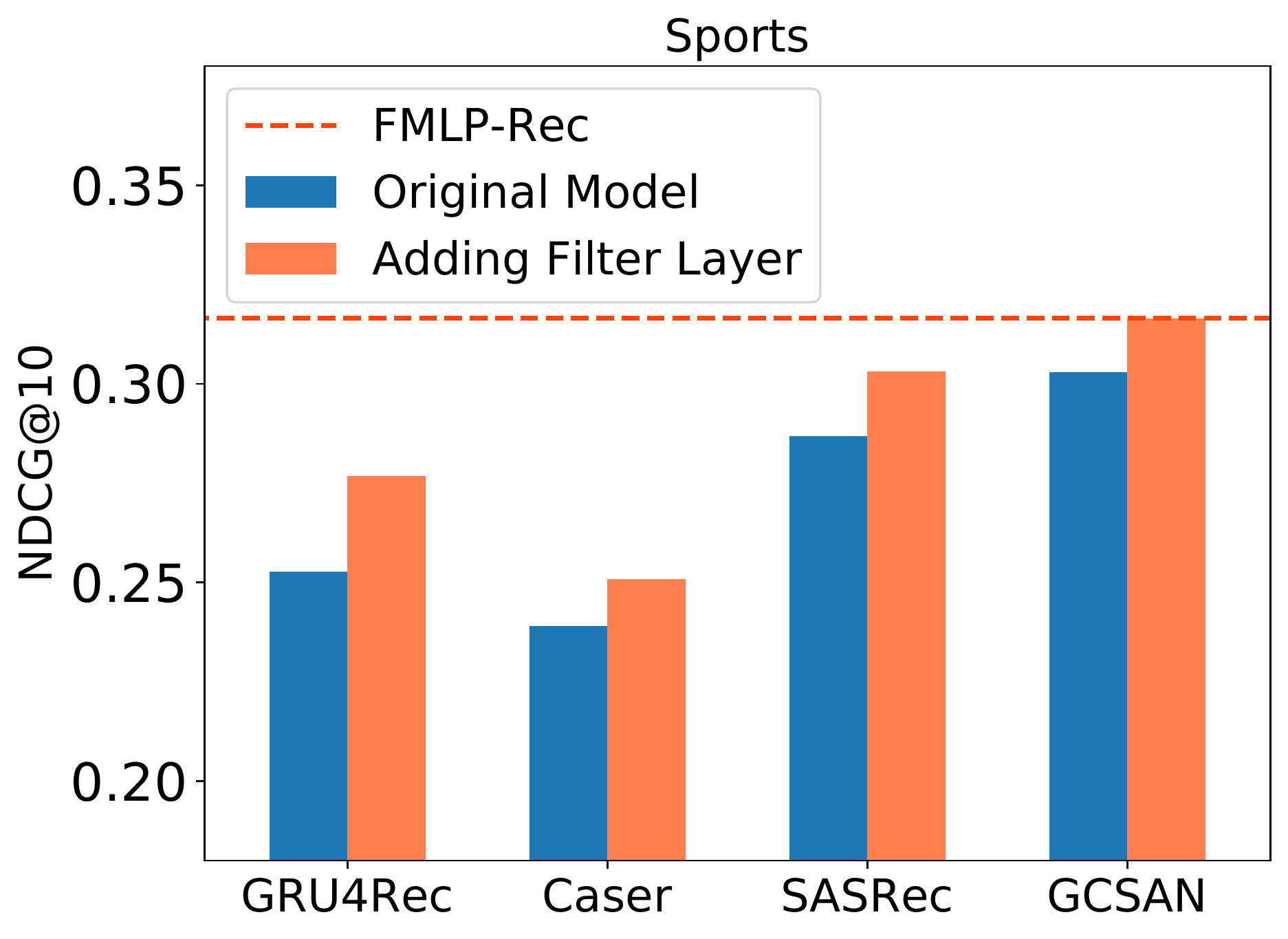}
        \label{toys-filter}
    \end{subfigure}
    \caption{Performance (NDCG@10) comparison of different models enhanced by our learnable filters.}
\label{fig-other-extend}
\end{figure}
The key contribution of our FMLP-Rec is the learnable filters. It is a general module that can be applied to  other sequential recommendation models. 
Thus, in this part, we examine whether our learnable filters can bring improvements to other models. 
Similar to the operations in Section~\ref{sec-motivation}, we add the learnable filters between the embedding layer and the sequence encoder layer, and select RNN-based GRU4Rec~\cite{DBLP:journals/corr/HidasiKBT15}, CNN-based Caser~\cite{DBLP:conf/wsdm/TangW18}, Transformer-based SASRec~\cite{DBLP:conf/icdm/KangM18} and GNN-based GCSAN~\cite{DBLP:conf/ijcai/XuZLSXZFZ19} as the base models.

The results are shown in Figure~\ref{fig-other-extend}. We also report the performance of FMLP-Rec for comparison.
First, after being integrated  with our learnable filters, all the baselines achieve better performance. It shows that the learnable filters are generally useful to reduce the influence of noise information for other models, even for different architectures.
Second, our FMLP-Rec still outperforms all the baselines and their variants. 
This is because our model only adopts MLP layers, which has less parameters and is more suitable for the learnable filters in sequential recommendation task.

\ignore{
\subsection{Visualization of Learned Filters}
\begin{figure}[t]
    \centering
    \begin{subfigure}[b]{0.49\linewidth}
        \centering
        \includegraphics[width=\textwidth]{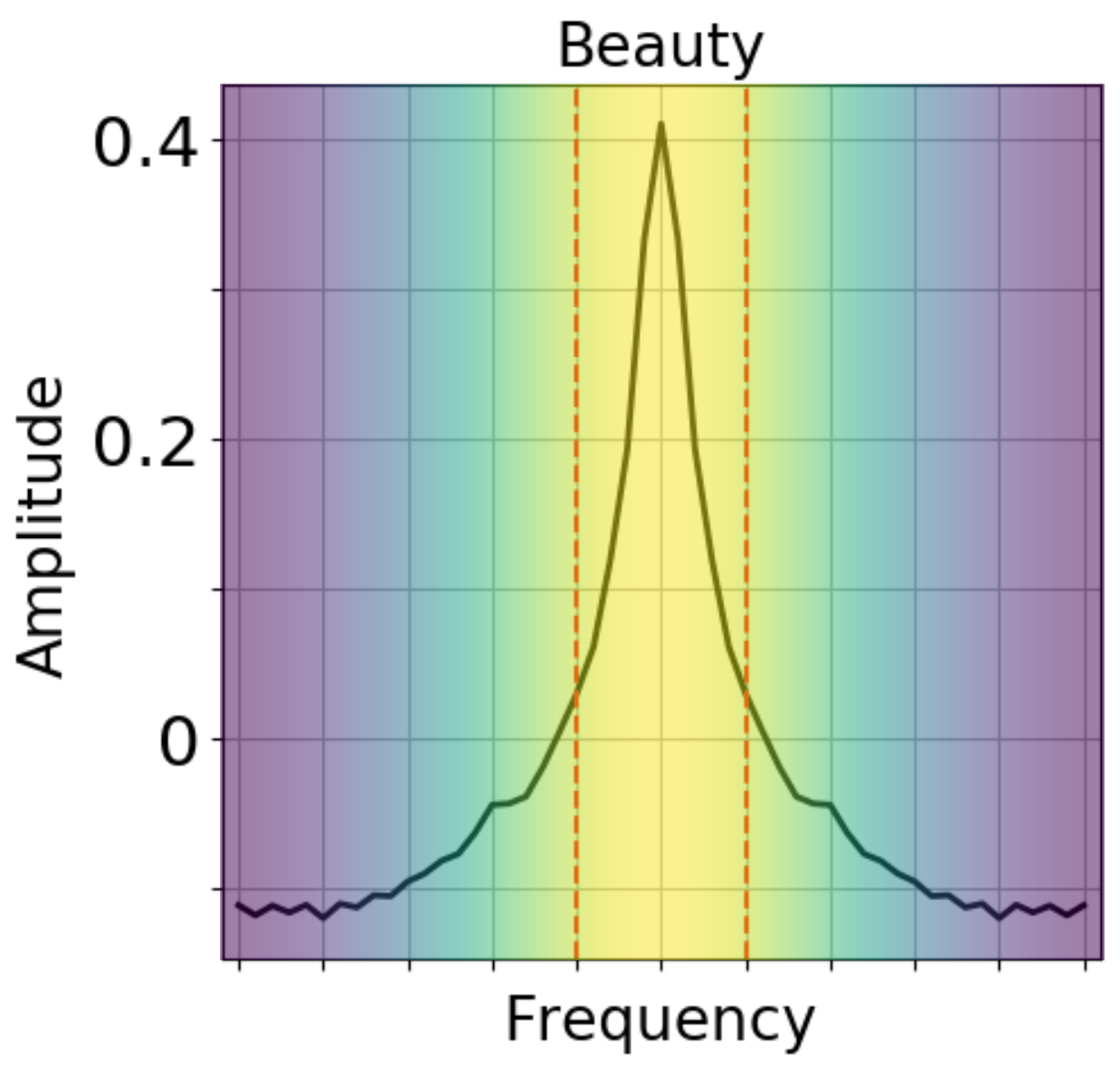}
        \label{beauty-conv-curve}
    \end{subfigure}
    \begin{subfigure}[b]{0.49\linewidth}
        \centering
        \includegraphics[width=\textwidth]{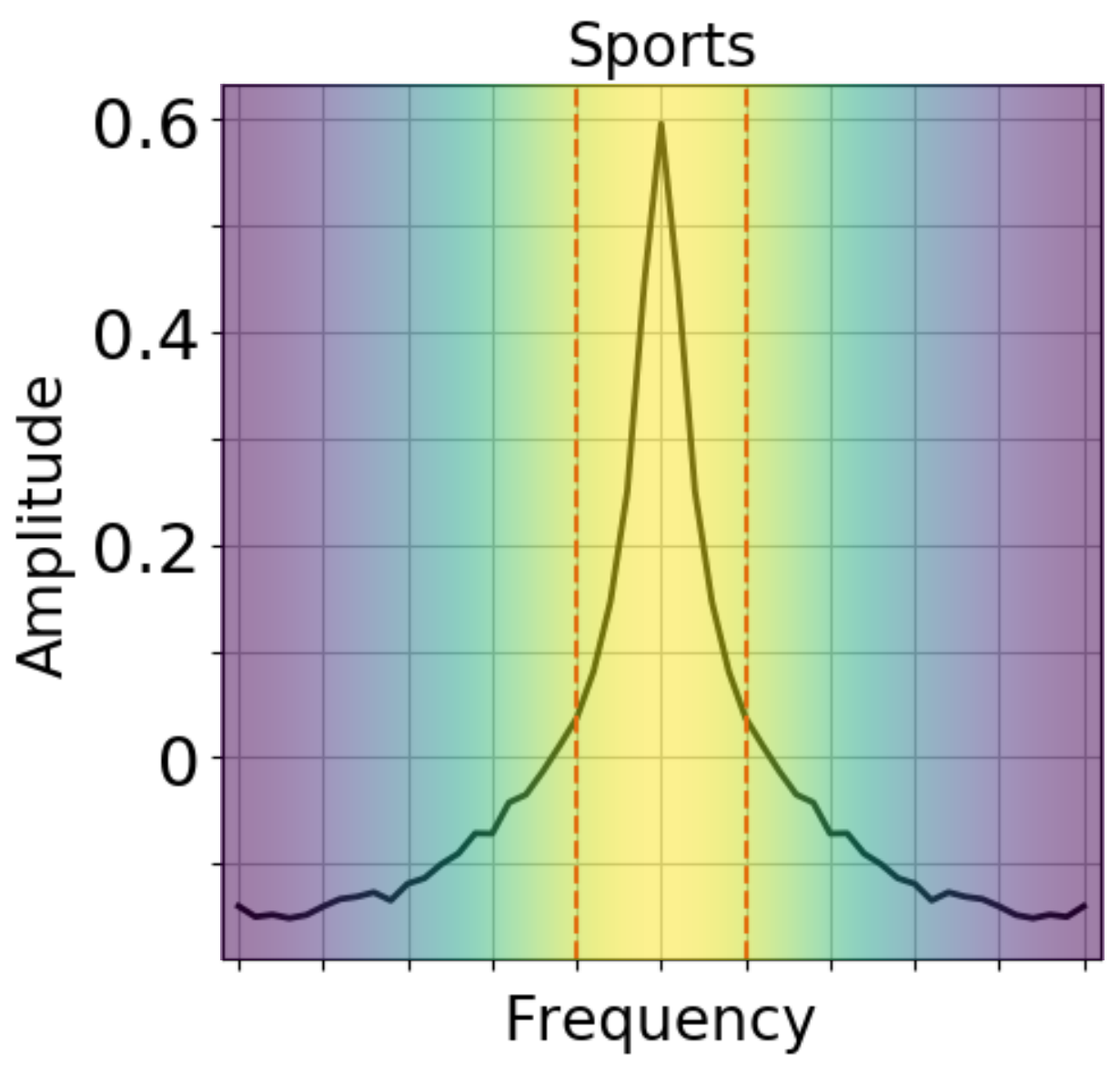}
        \label{toys-conv-curve}
    \end{subfigure}
    \caption{Visualization of our learned filters, the amplitude denotes the weight from $\bX$ to multiple the corresponding frequency in $\bF^{l}$.}
\label{fig-filter-curve}
\end{figure}
The core operation in FMLP-Rec is the the element-wise multiplication between frequency domain features $\bF^{l}$ and the learnable filter $\bX$ in Eq.~\ref{eq-multi-fre}. 
In this part, we visualize and interpret the learnable filters in the frequency domain.
We average the values of learnable filters in the first layer for all dimensions of features, and depict it in Figure~\ref{fig-filter-curve}.

We can see that the learnable filters have clear patterns in the frequency domain, where the low-frequency signals are assigned larger positive weights than high-frequency signals.
Therefore, the learnable filters can be seen as special low-pass filters for attenuate the high-frequency noise.
This finding is similar to the empirical results in Section~\ref{sec-motivation} that high-frequency information in item embedding matrix is usually noise.}
\section{RELATED WORK}

\paratitle{Sequential Recommendation.}
Early works~\cite{DBLP:conf/www/RendleFS10} on sequential recommendation are based on the Markov Chain assumption and focus on modeling item-item transition relationships to predict the next item given the last interaction of a user.
A series of works follow this line and extend it to high-order MCs~\cite{DBLP:conf/wsdm/TangW18,DBLP:conf/icdm/KangM18,DBLP:conf/recsys/HidasiQKT16}.
With the development of the neural networks, Hidasi et al.~\cite{DBLP:journals/corr/HidasiKBT15} introduced GRU to capture sequential patterns, and a surge of works leverage other neural network architectures for sequential recommendation, \eg CNN~\cite{DBLP:conf/wsdm/TangW18}, GNN~\cite{DBLP:conf/ijcai/XuZLSXZFZ19,tan2021sparse} and Transformer~\cite{DBLP:conf/icdm/KangM18,DBLP:conf/cikm/SunLWPLOJ19}.
Based on these neural network architectures, various studies introduce other contextual information (\eg item attributes and reviews) by adding memory networks~\cite{DBLP:conf/sigir/HuangZDWC18}, hierarchical structures~\cite{DBLP:conf/ijcai/LiNLCQ19}, data augmentation~\cite{zhang2021causerec,wang2021counterfactual} and pre-training technique~\cite{DBLP:conf/cikm/ZhouWZZWZWW20,DBLP:conf/www/BianZZCCHLW21,DBLP:conf/cikm/BianZZCHYW21}, etc.
Despite the success of these deep models in the sequential recommendation task, we find that these models are easy to be affected by noise information from the user historical behaviors.
To solve this problem, we adopt learnable filters to reduce the influence of noise signals, and devise a lightweight all-MLP architecture to alleviate overfitting.

\paratitle{All-MLP Models.}
Multi-Layer Perceptron (MLP)~\cite{DBLP:conf/nips/YairG88} is a classical feed-forward neural network that consists of fully-connected neurons and non-linear activation functions.
It is widely used as assistant modules to couple with CNN~\cite{DBLP:conf/nips/BengioCH93}, RNN~\cite{DBLP:conf/nips/Pineda87} and self-attention network~\cite{DBLP:conf/nips/VaswaniSPUJGKP17} to construct deep models.
Recently, several studies question the necessity of CNN~\cite{DBLP:journals/corr/abs-2105-01601} and self-attention network~\cite{DBLP:journals/corr/abs-2107-08391,DBLP:journals/corr/abs-2104-05707}, and propose to use MLP to replace the above architectures.
These all-MLP models mostly aim to devise effective MLP-based mixing architectures to capture the interaction of input information, \eg mixer layer~\cite{DBLP:journals/corr/abs-2105-01601}, axial shift block~\cite{DBLP:journals/corr/abs-2107-08391} and spatial shift block~\cite{DBLP:journals/corr/abs-2105-08050}, and have performed well in various tasks, \eg image classification~\cite{DBLP:journals/corr/abs-2105-01601}and semantic segmentation~\cite{DBLP:journals/corr/abs-2107-08391}.
More recently, GFNet~\cite{DBLP:journals/corr/abs-2107-00645} adopts 2D Fourier transform with the global filter layer to learn long-term spatial dependencies in the frequency domain, which shows competitive performance to Transformer-based models.
However, all-MLP models are hard to capture sequential characteristics, which are essential to the sequential recommendation task.
In our approach, we design an all-MLP model that includes the filter layers to encode the sequence in the frequency domain.
It outperforms competitive RNN, CNN and Transformer-based baseline models in eight datasets, with a lightweight architecture.
\section{Conclusion}
In this paper, we found that the logged user behavior data usually  contains noisy interactions, and performed empirical study analysis to show that filtering algorithms from the digital signal processing area are useful to alleviate the influence of the noise in deep sequential recommendation models.
Inspired by it, we proposed FMLP-Rec, an all-MLP model with learnable filters for sequential recommendation task.
The all-MLP architectures endowed our model with lower time complexity, and the learnable filters can be optimized by SGD to adaptively attenuate the noise information in the frequency domain.
We also showed that the learnable filters are equivalent to the circular convolution in the time domain, which have a larger receptive field and can better capture periodic characteristics.
Experimental results have shown that our approach outperforms several competitive RNN, CNN, GNN and Transformer-based baselines. 

\begin{acks}
This work was partially supported by the National Natural Science Foundation of China under Grant No. 61872369 and 61832017, Beijing Outstanding Young Scientist Program under Grant No. BJJWZYJH012019100020098, 
the Outstanding Innovative Talents Cultivation Funded Programs 2021 and Public Computing Cloud, Renmin University of China. This work is supported by Beijing Academy of Artificial Intelligence~(BAAI). Xin Zhao is the corresponding author.
\end{acks}
\bibliographystyle{ACM-Reference-Format}
\bibliography{sample-base}

\newpage
\appendix
\ignore{
\begin{figure*}[htb]
    \centering
    \begin{subfigure}[b]{0.24\linewidth}
        \centering
        \includegraphics[width=\textwidth]{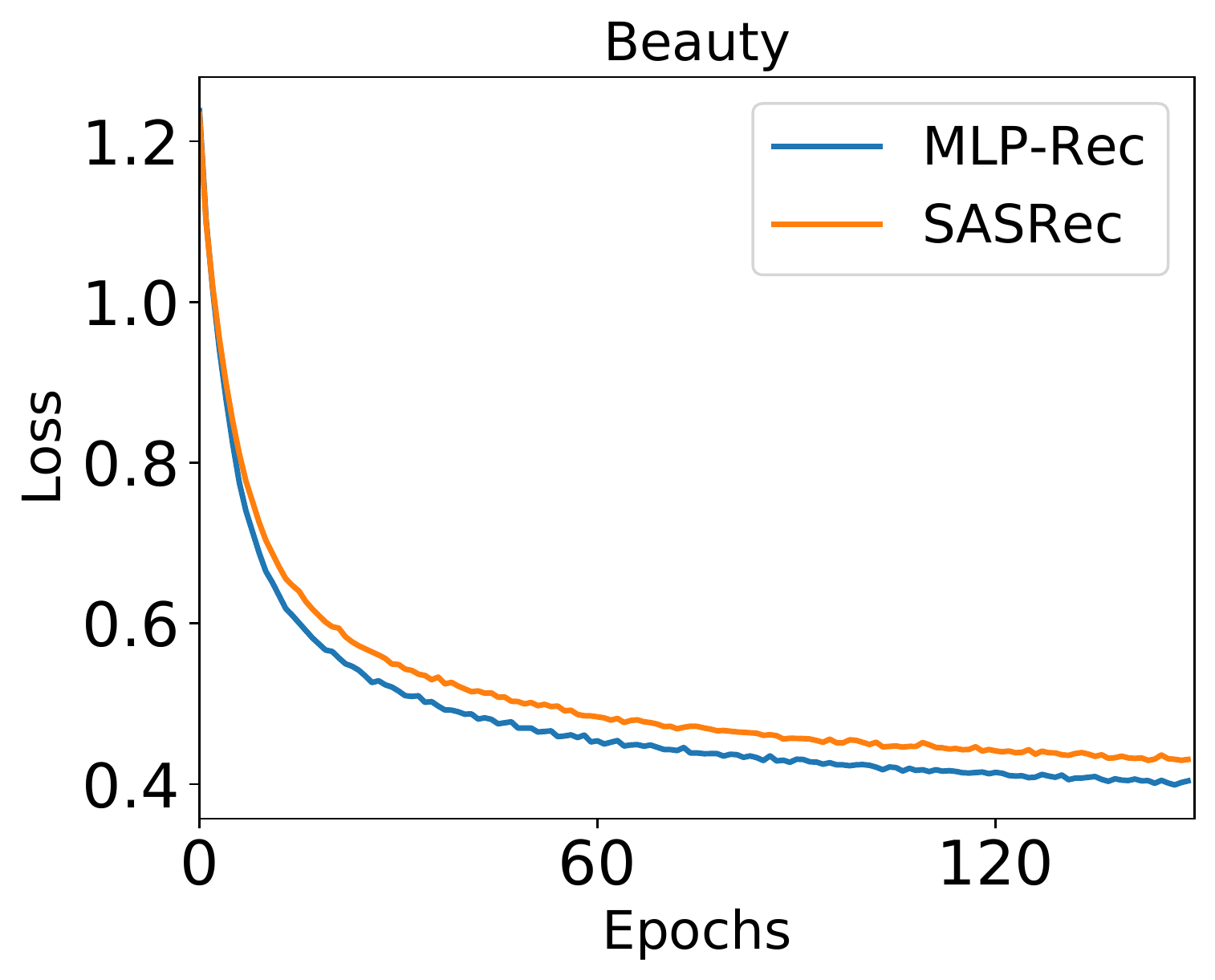}
    \end{subfigure}
    \begin{subfigure}[b]{0.24\linewidth}
        \centering
        \includegraphics[width=\textwidth]{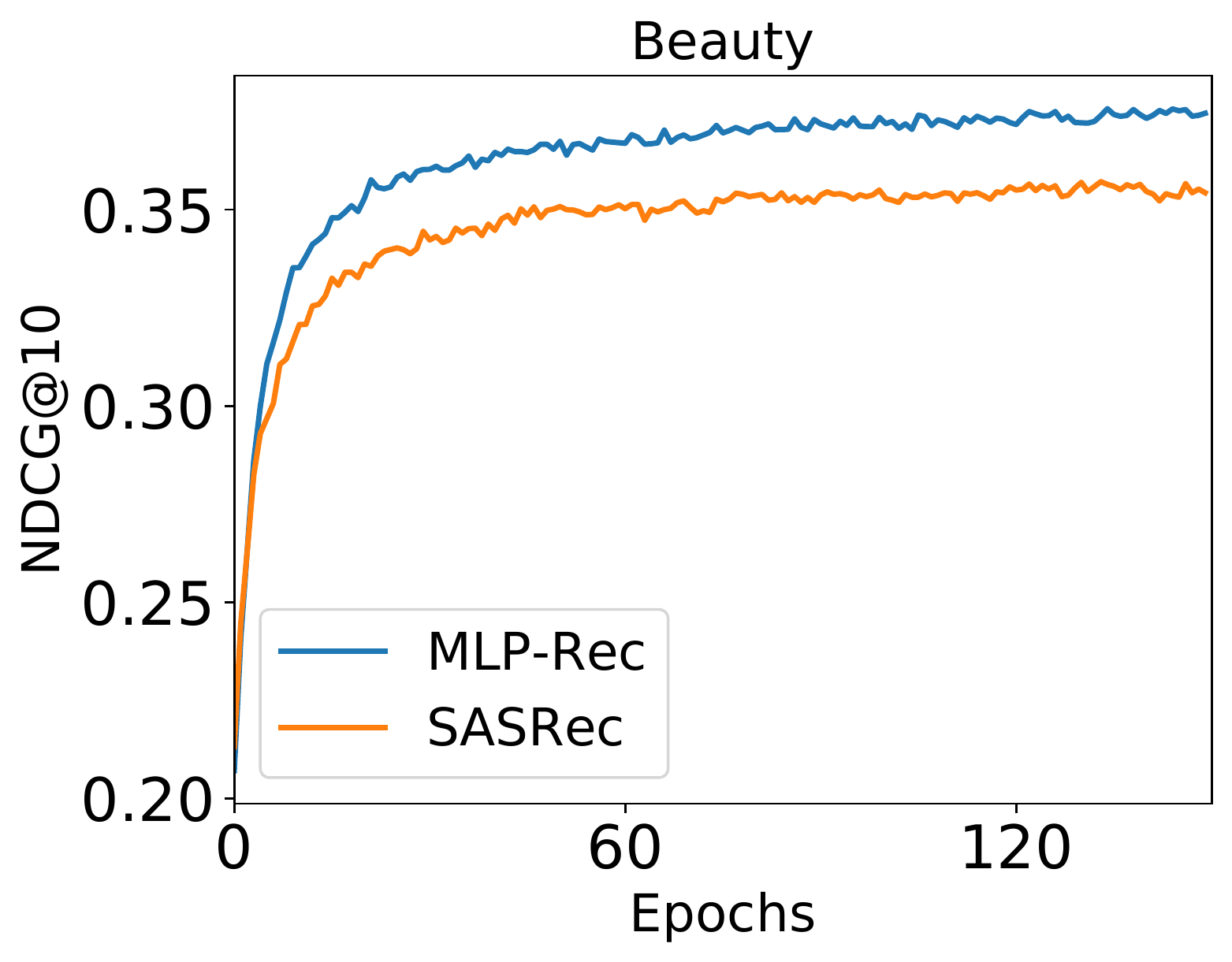}
    \end{subfigure}
    \begin{subfigure}[b]{0.24\linewidth}
        \centering
        \includegraphics[width=\textwidth]{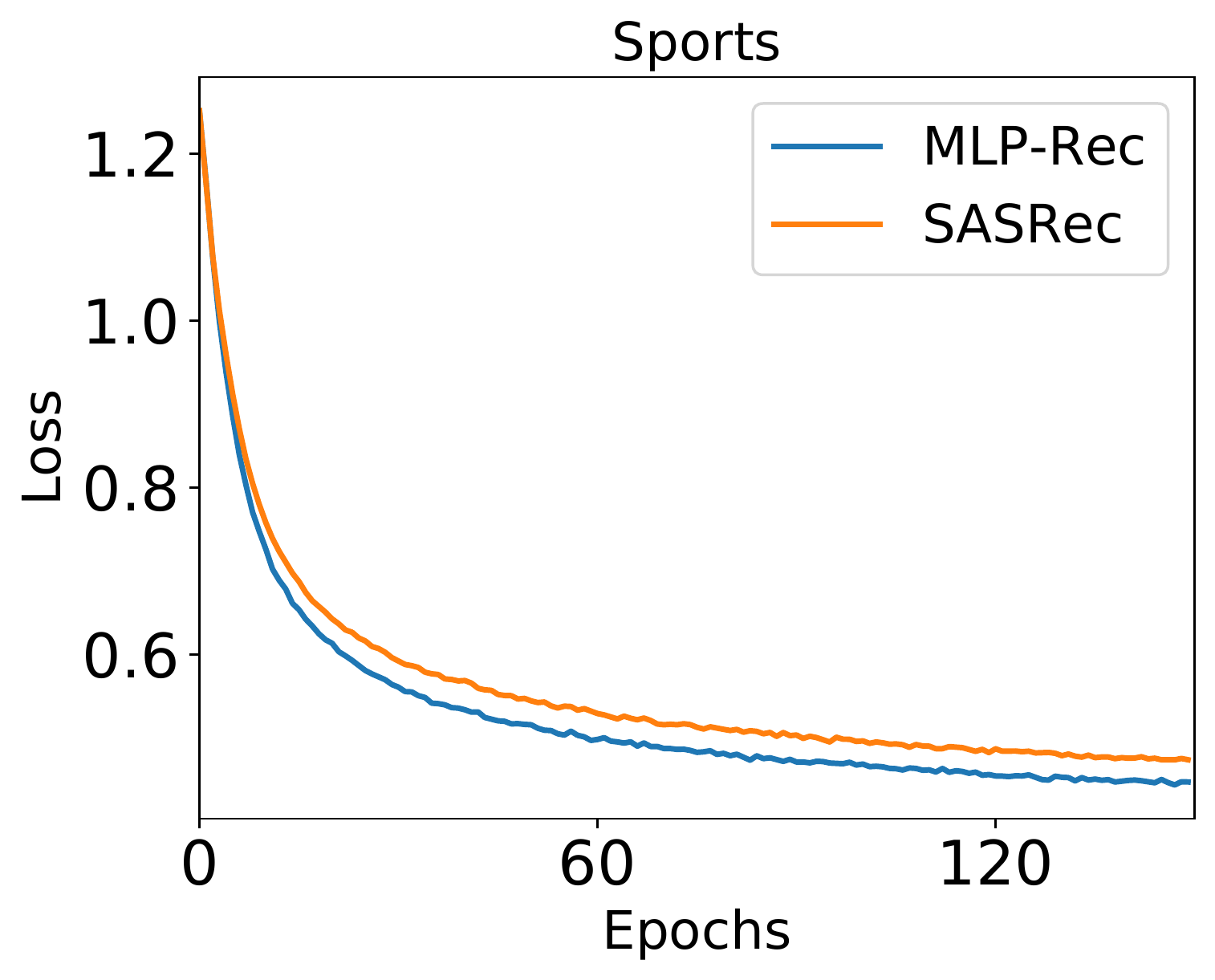}
    \end{subfigure}
    \begin{subfigure}[b]{0.24\linewidth}
        \centering
        \includegraphics[width=\textwidth]{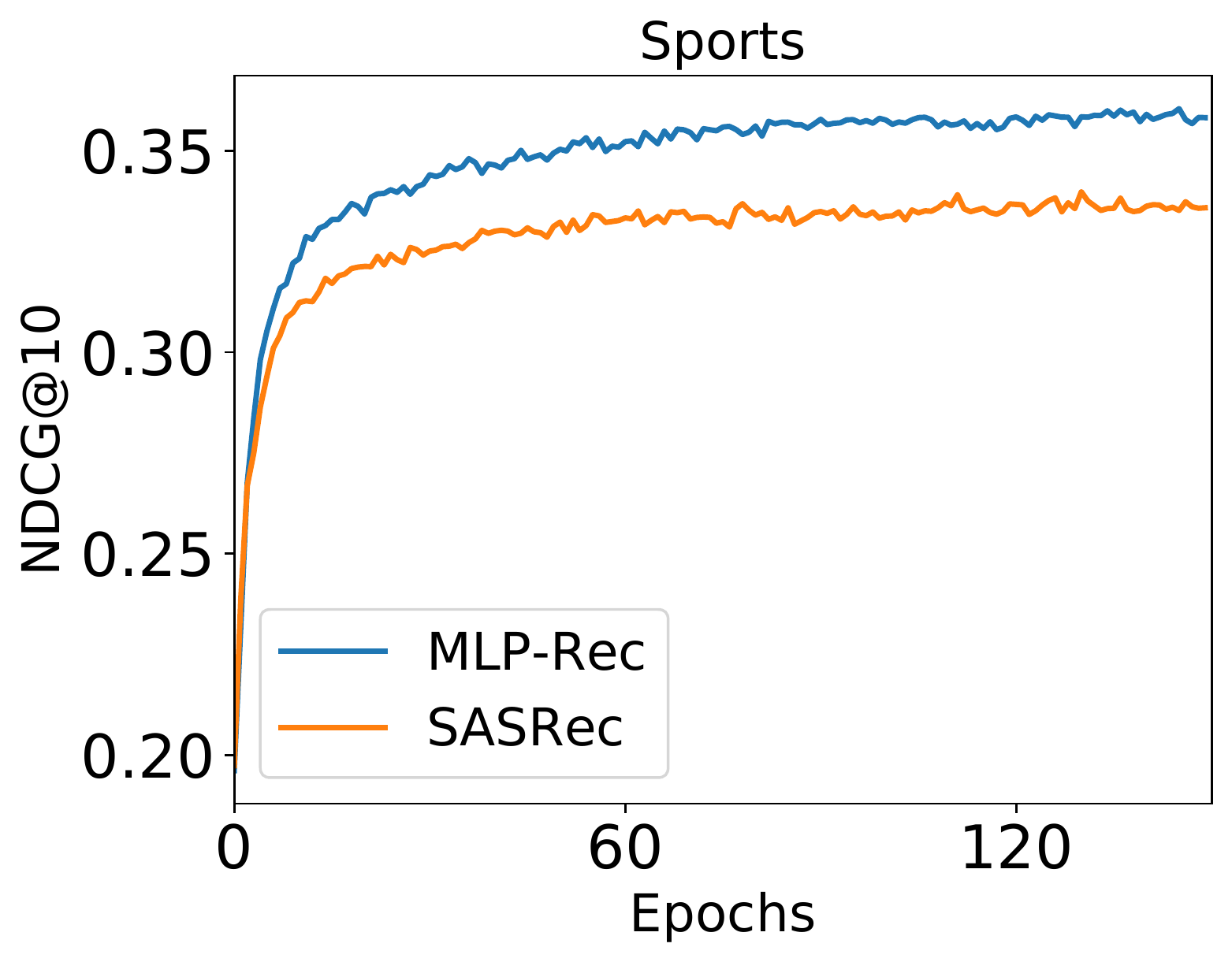}
    \end{subfigure}
    
    \centering
    \begin{subfigure}[b]{0.24\linewidth}
        \centering
        \includegraphics[width=\textwidth]{v4/images/Beauty_loss.pdf}
    \end{subfigure}
    \begin{subfigure}[b]{0.24\linewidth}
        \centering
        \includegraphics[width=\textwidth]{v4/images/Beauty_ndcg.pdf}
    \end{subfigure}
    \begin{subfigure}[b]{0.24\linewidth}
        \centering
        \includegraphics[width=\textwidth]{v4/images/Sports_loss.pdf}
    \end{subfigure}
    \begin{subfigure}[b]{0.24\linewidth}
        \centering
        \includegraphics[width=\textwidth]{v4/images/Sports_ndcg.pdf}
    \end{subfigure}
    \caption{Training loss and Testing recall (NDCG@10) of our approach and SASRec on Beauty and Sports datasets with the increasing epochs.}
\label{fig-training}
\end{figure*}}

\section{Implementation Details}
For Caser, HGN and BERT4Rec, we use the source code provided by their authors. 
For CLEA, we implement it by PyTorch~\footnote{https://pytorch.org/}.
For other methods, we implement them based on RecBole~\cite{DBLP:journals/corr/abs-2011-01731}. 
All hyper-parameters are set following the suggestions from the original papers.
For our proposed FMLP-Rec, we develop it based on the codes and data released by S$^3$-Rec~\cite{DBLP:conf/cikm/ZhouWZZWZWW20}, the dimension of the embedding is 64, and the maximum sequence length is 50.
We set the number of the learnable filter blocks as 2, the batch size is set as 256.
We use the Adam optimizer~\cite{DBLP:journals/corr/KingmaB14} with a learning rate of 0.001, and adopt early-stopped training if the MRR performance on the validation set decreases for 10 continuous epochs.

\section{More Analysis}
\subsection{Training Curves Analysis}
\begin{figure}
    \centering
    \begin{subfigure}[b]{0.49\linewidth}
        \centering
        \includegraphics[width=\textwidth]{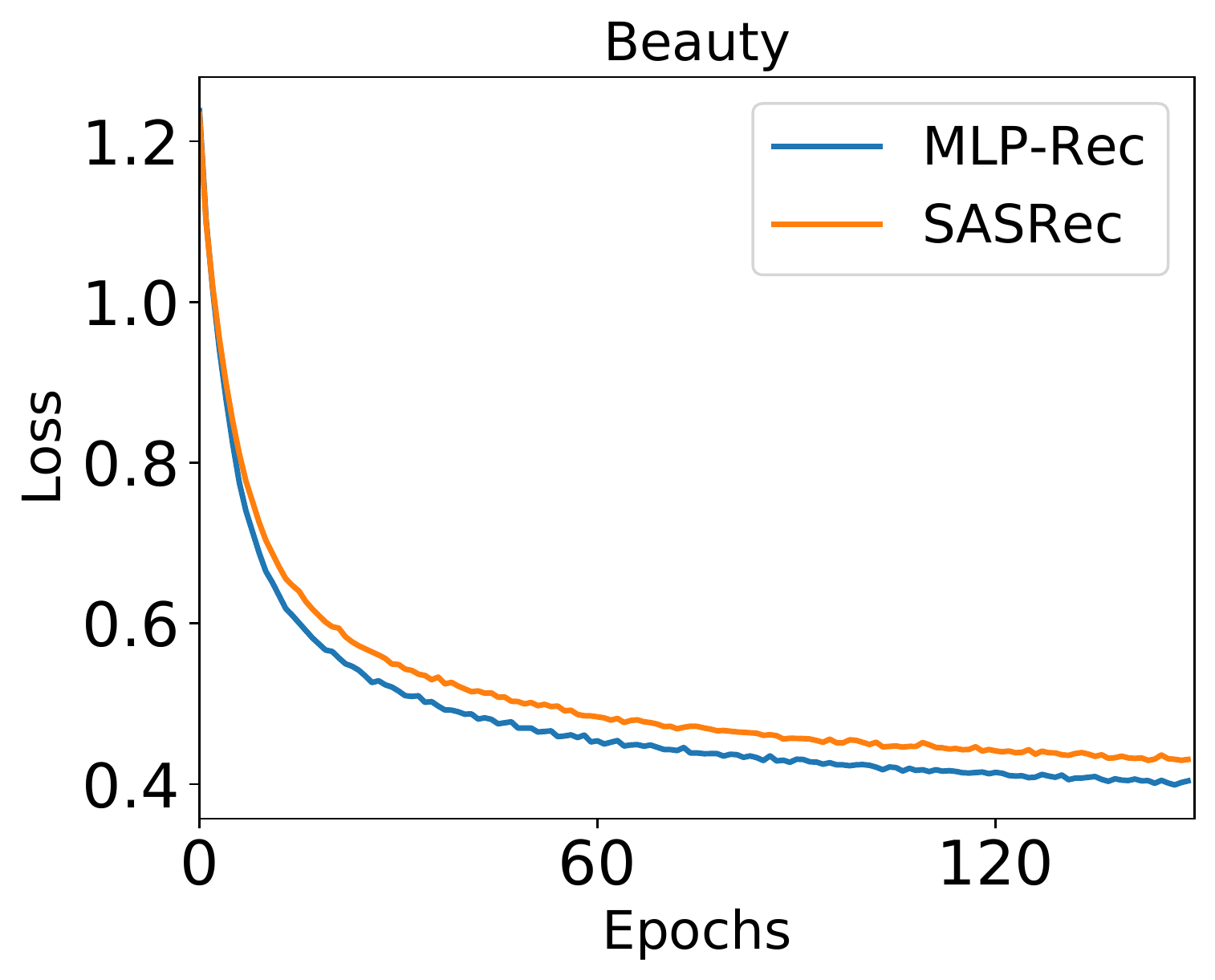}
    \end{subfigure}
    \begin{subfigure}[b]{0.49\linewidth}
        \centering
        \includegraphics[width=\textwidth]{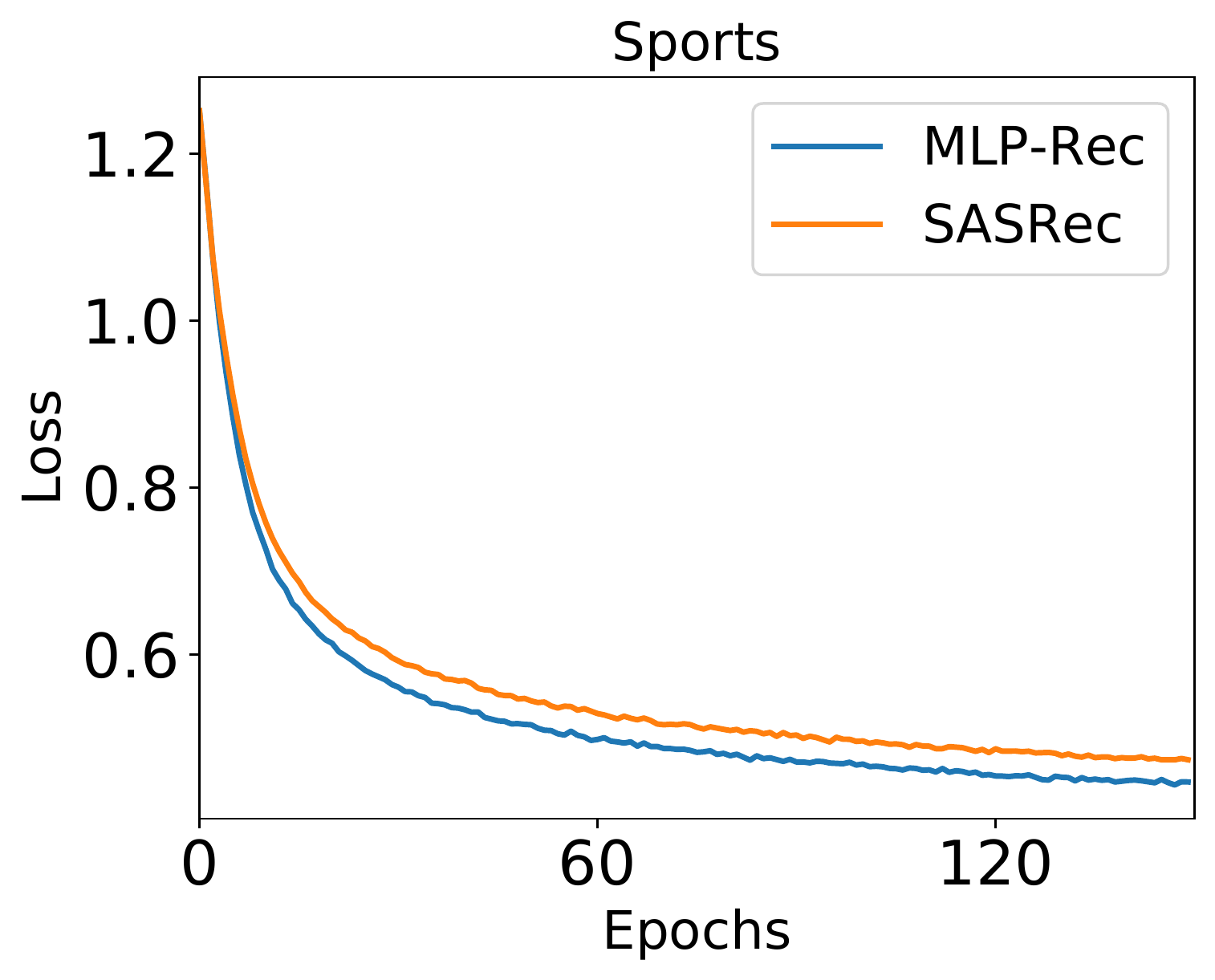}
    \end{subfigure}
    \caption{Training loss of our approach and SASRec on Beauty and Sports datasets with the increasing epochs.}
\label{fig-training-loss}
\end{figure}

\begin{figure}
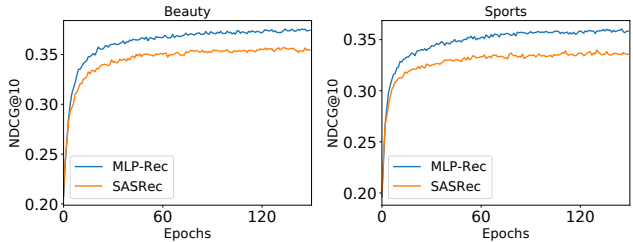

    \centering
    \begin{subfigure}[b]{0.49\linewidth}
        \centering
        \includegraphics[width=\textwidth]{images/Beauty_ndcg.pdf}
    \end{subfigure}
    \begin{subfigure}[b]{0.49\linewidth}
        \centering
        \includegraphics[width=\textwidth]{images/Sports_ndcg.pdf}
    \end{subfigure}
    \caption{Testing recall (NDCG@10) of our approach and SASRec on Beauty and Sports datasets with the increasing epochs.}
\label{fig-training-recall}
\end{figure}
To reveal that our proposed FMLP-Rec is able to converge better than baseline models, we analyze the training process of SASRec and FMLP-Rec.
We conduct the experiments on Beauty and Sports datasets and plot the curves of training loss and testing recall of the two models in Figure~\ref{fig-training-loss} and Figure~\ref{fig-training-recall}, respectively.
Note that in the figures we show the training processes under the optimal hyper-parameter setting for both methods.

In these figures, we can find that along the training process, FMLP-Rec consistently obtains lower training loss, which indicates that FMLP-Rec is able to better fit the training data than SASRec.
Moreover, the lower training loss successfully transfers to better testing accuracy, which indicates the strong generalization capacity of our FMLP-Rec.
The reason is that FMLP-Rec adopts learnable filters to attenuate noise information within the item representation matrices, which can better capture the useful characteristics and alleviate the overfitting problem on noise information.
In contrast, the higher training loss and lower testing accuracy of SASRec reflect the practical difficulty to train an over-parameterized Transformer-based model well.

\subsection{Parameter Tuning}
\begin{figure}
    \centering
    \begin{subfigure}[b]{0.49\linewidth}
        \centering
        \includegraphics[width=\textwidth]{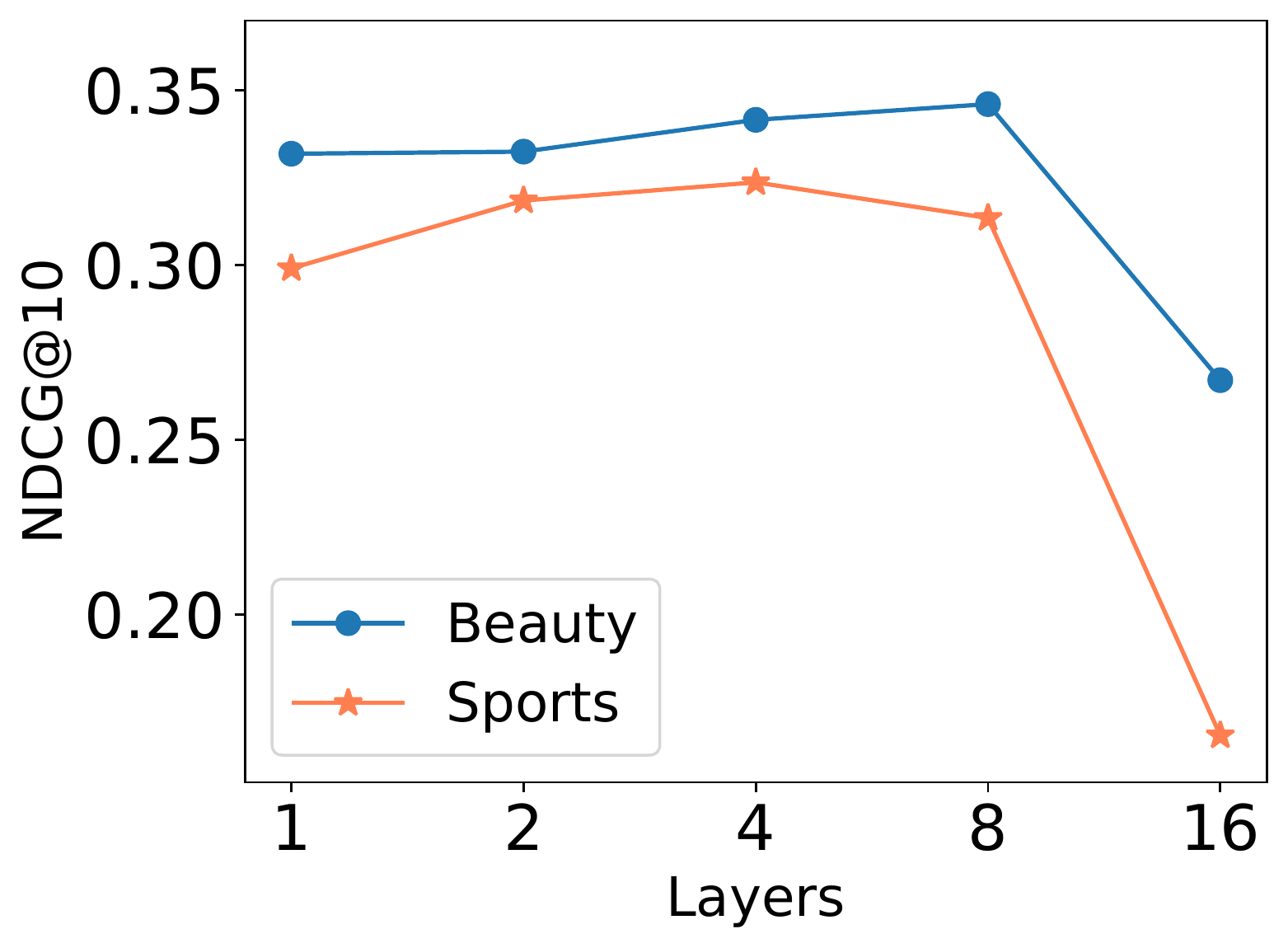}
    \end{subfigure}
    \begin{subfigure}[b]{0.49\linewidth}
        \centering
        \includegraphics[width=\textwidth]{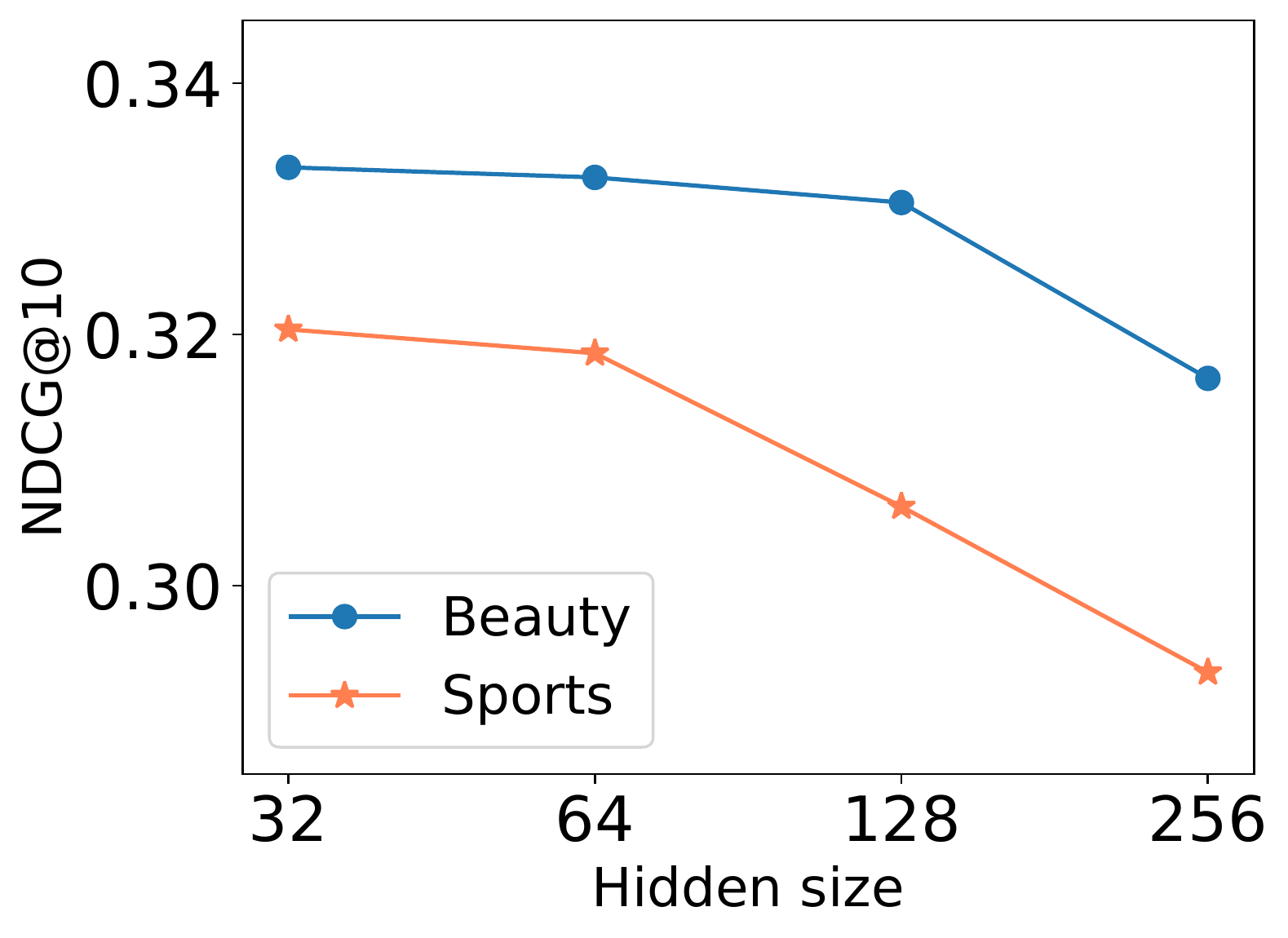}
    \end{subfigure}
    \caption{Performance (NDCG@10) comparison w.r.t. different numbers of pre-training epochs on Beauty and Toys datasets.}
\label{fig-hyperparameter}
\end{figure}

When applying our FMLP-Rec to a new dataset, besides the standard hyper-parameters \eg learning rate and training epochs, the most important hyper-parameters to tune are the layer number and the hidden size. 
Here we investigate the performance change of FMLP-Rec with the respect to the layer number and the hidden size on Beauty and Sports datasets.

As shown in Figure~\ref{fig-hyperparameter}, with the increasing of the layer number, the performance of FMLP-Rec can be further improved.
However, when the layer number achieves 16, the gain starts to become trivial.
It indicates that it is promising to deepen our FMLP-Rec for achieving more exciting performance, but the most proper architecture requires to be further investigated.
We leave it into future works.
For hidden size, we can see that the best value is about 64. 
The reason may be that too large hidden size can involve more parameters that are easy to overfit into the noise information, while too small size is not enough to capture the user preference. 

\subsection{Visualization of Learned Filters on Other Datasets}
\begin{figure*}
    \centering
    \begin{subfigure}[b]{0.24\linewidth}
        \centering
        \includegraphics[width=\textwidth]{images/Filter_Beauty.pdf}
    \end{subfigure}
    \begin{subfigure}[b]{0.24\linewidth}
        \centering
        \includegraphics[width=\textwidth]{images/Filter_Sports.pdf}
    \end{subfigure}
    \begin{subfigure}[b]{0.24\linewidth}
        \centering
        \includegraphics[width=\textwidth]{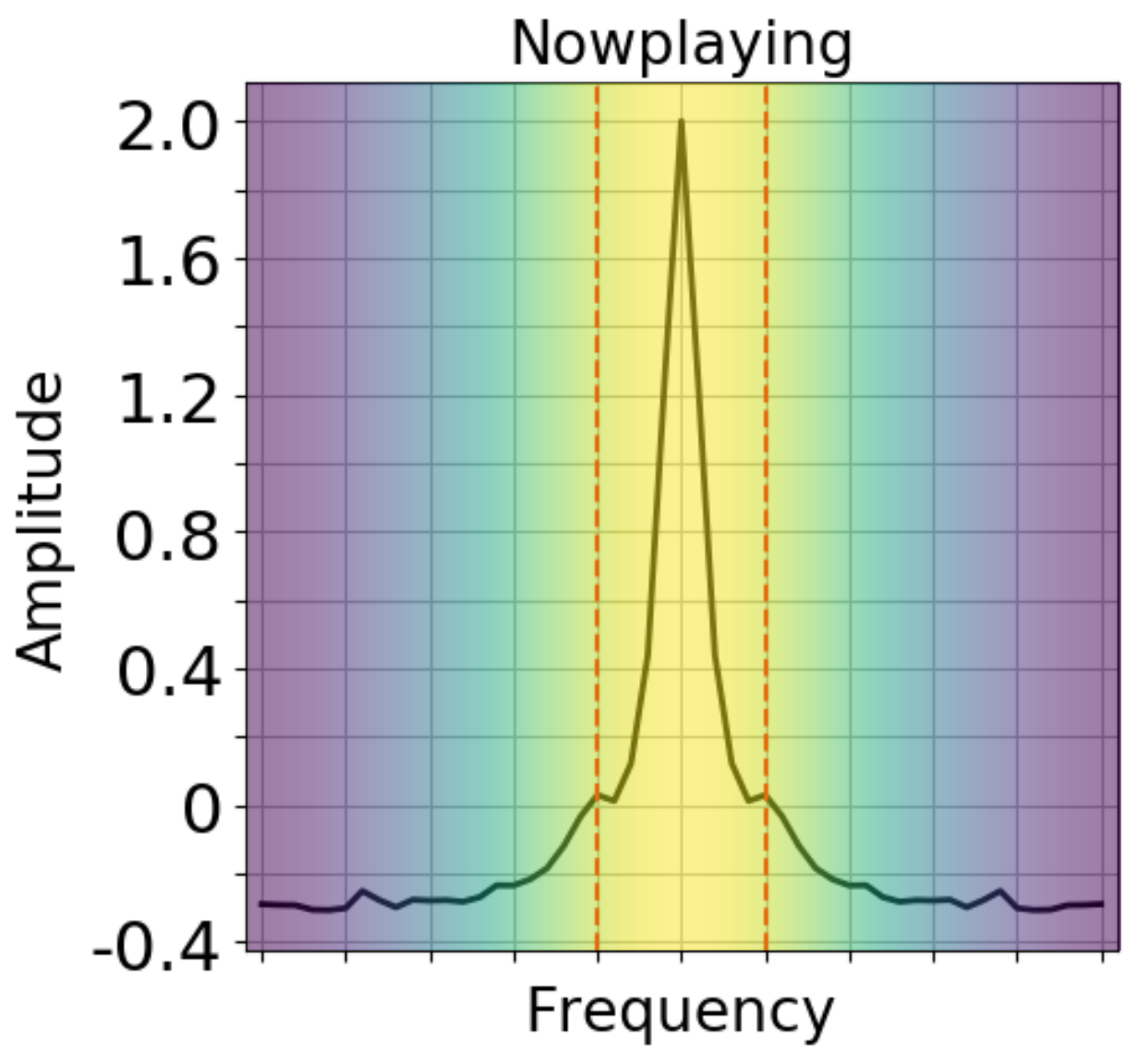}
    \end{subfigure}
    \begin{subfigure}[b]{0.24\linewidth}
        \centering
        \includegraphics[width=\textwidth]{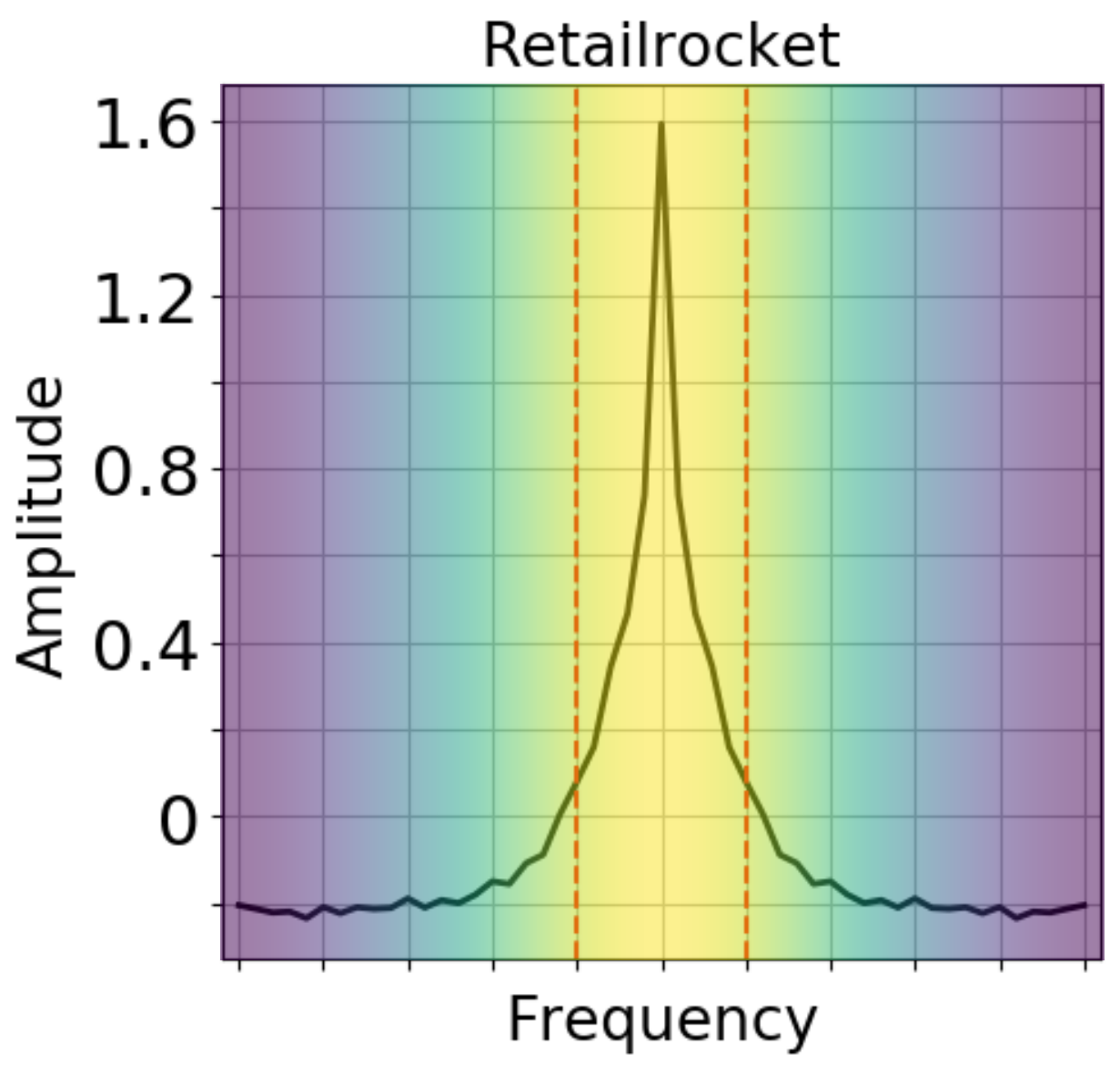}
    \end{subfigure}
    \caption{Visualization of our learned filters on four datasets, the amplitude denotes the weight from $\bX$ to multiply the corresponding frequency in $\bF^{l}$.}
\label{fig-visual-other}
\end{figure*}

The core operation in FMLP-Rec is the the element-wise multiplication between frequency domain features $\bF^{l}$ and the learnable filter $\bX$ in Eq.~\ref{eq-multi-fre}. 
In this part, we visualize and interpret the learnable filters in the frequency domain.
We average the values of learnable filters in the first layer for all dimensions of features, and depict it in Figure~\ref{fig-visual-other}.

We can see that the learnable filters have clear patterns in the frequency domain, where the low-frequency signals are assigned larger positive weights than high-frequency signals.
Therefore, the learnable filters can be seen as special low-pass filters to attenuate the high-frequency noise.
This finding is similar to the empirical results in Section~\ref{sec-motivation} that high-frequency information in item embedding matrix is usually noise.

\subsection{Performance under Full-Ranking Setting}
\begin{table}[t!]
    \small
	\caption{Performance comparison of different methods on four datasets under full-sort setting. The best performance and the second best performance methods are denoted in bold and underlined fonts respectively.}
	\label{tab:side_table}
	\setlength{\tabcolsep}{1.3mm}{
	\begin{tabular}{llcccc}
	\toprule
		Datasets & Metric  & GRU4Rec & Caser  &SASRec  &FMLP-Rec  \\
	\midrule
\multirow{6} * {Beauty}
 &HR@5   &0.0164  &0.0205    &\underline{0.0387}  &\textbf{0.0398} \\
&NDCG@5   &0.0099  &0.0131    &\underline{0.0249}   &\textbf{0.0258} \\
&HR@10   &0.0283  &0.0347    &\underline{0.0605}   &\textbf{0.0632} \\
&NDCG@10   &0.0137  &0.0176    &\underline{0.0318}   &\textbf{0.0333} \\
&HR@20   &0.0479  &0.0556    &\underline{0.0902}   &\textbf{0.0958} \\
&NDCG@20   &0.0187  &0.0229    &\underline{0.0394}   &\textbf{0.0415} \\
\hline
\multirow{6} * {Sports}
&HR@5   &0.0129  &0.0116    &\textbf{0.0233}   &\underline{0.0218} \\
&NDCG@5   &0.0086  &0.0072    &\textbf{0.0154}   &\underline{0.0144} \\
&HR@10   &0.0204  &0.0194    &\textbf{0.0350}   &\underline{0.0344} \\
&NDCG@10   &0.0110  &0.0097    &\textbf{0.0192}   &\underline{0.0185} \\
&HR@20   &0.0333  &0.0314    &\underline{0.0507}   &\textbf{0.0537} \\
&NDCG@20   &0.0142  &0.0126    &\underline{0.0231}   &\textbf{0.0233} \\
\hline
\multirow{6} * {Toys}
 &HR@5   &0.0097  &0.0166    &\textbf{0.0463}   &\underline{0.0456} \\
&NDCG@5   &0.0059  &0.0107    &\underline{0.0306}   &\textbf{0.0317} \\
&HR@10   &0.0176  &0.0270    &\underline{0.0675}   &\textbf{0.0683} \\
&NDCG@10   &0.0084  &0.0141    &\underline{0.0374}   &\textbf{0.0391} \\
&HR@20   &0.0301  &0.0420    &\underline{0.0941}   &\textbf{0.0991} \\
&NDCG@20   &0.0116  &0.0179   &\underline{0.0441}   &\textbf{0.0468} \\
\hline
\multirow{6} * {Yelp}
 &HR@5   &0.0152  &0.0151    &\underline{0.0162}  &\textbf{0.0179} \\
&NDCG@5   &0.0099  &0.0096    &\underline{0.0100}   &\textbf{0.0113} \\
&HR@10   &0.0263  &0.0253    &\underline{0.0274}   &\textbf{0.0304} \\
&NDCG@10   &0.0134  &0.0129    &\underline{0.0136}   &\textbf{0.0153} \\
&HR@20   &0.0439  &0.0422    &\underline{0.0457}   &\textbf{0.0511} \\
&NDCG@20   &0.0178  &0.0171    &\underline{0.0182}   &\textbf{0.0205} \\

\bottomrule
\end{tabular}}
\end{table}
To further verify the effectiveness of our FMLP-Rec, we also conduct experiments under the full-ranking setting.
We select RNN-based GRU4Rec, CNN-based Caser, and Transformer-based SASRec as representative baseline models. The results on Beauty, Sports, Toys and Yelp datasets are shown in Table~\ref{tab:side_table}.

From the table, we can see similar tendency as in Table~\ref{tab:main_table_cikm2020}.
It can be found that SASRec and FMLP-Rec perform much better than the other three methods.
In most of cases, FMLP-Rec outperforms SASRec a lot. It indicates that the all-MLP architecture and learnable filters are exactly effective for the sequential recommendation task.


\end{document}